\documentclass[10pt]{article}

\usepackage{blindtext}
\usepackage[a4paper, total={6in, 8in}]{geometry}

\usepackage{booktabs} 
\usepackage{orcidlink}
\usepackage[english]{babel}
\usepackage{wrapfig,booktabs}
\usepackage[utf8x]{inputenc}
\usepackage{graphicx}
\usepackage[colorinlistoftodos]{todonotes}
\usepackage{amsmath}
\usepackage{amssymb}
\usepackage{xcolor}
\usepackage{framed} 
\usepackage{colortbl}
\usepackage{url}
\usepackage[noend]{algorithmic}
\usepackage{array}
\usepackage{etoolbox}
\usepackage{listings}
\usepackage{pgfplots}
\usepackage{multicol}
\usepackage{adjustbox}
\usepackage{subcaption}
\usepackage[inline]{enumitem}
\usepackage[binary-units]{siunitx}

\usepgfplotslibrary{units}

\def\ie{\textit{i.e., }}
\def\eg{\textit{e.g.}}
\def\wrt{with respect to}

\usepackage[linesnumbered,ruled]{algorithm2e}

\usepackage{todonotes}

\newtheorem{example}{Example}
\newtheorem{definition}{Definition}
\newtheorem{proposition}{Proposition}

\newtheorem{corollary}{Corollary}

\newcommand{\X}{\hbox{\textbf{X}}\xspace}

\def\sqr#1#2{{\vcenter{\vbox{\hrule height.#2pt \hbox{\vrule width.#2pt height#1pt
					\kern#1pt \vrule width.#2pt}
				\hrule height.#2pt}}}}

\newcommand{\Ct}{\hbox{$\mathbb {C}$}}
\newcommand{\Db}{\hbox{$\mathfrak {D}$}}

\newcommand{\del}{\hbox{\textsf{dRequest}}\xspace}
\newcommand{\delIso}{\hbox{\textsf{isoDel}}\xspace}
\newcommand{\ins}{\hbox{\textsf{iRequest}}\xspace}

\begin{document}

\title{Incremental Consistent Updating of  Incomplete  Databases\\
(Extended Version - Technical Report)}

\author{Jacques Chabin \orcidlink{0000-0003-1460-9979}, Mirian Halfeld Ferrari \orcidlink{0000-0003-2601-3224}\\
LIFO -- Universit\'e d'Orl\'eans, INSA CVL -- Orl\'eans, France
\\~\\
Nicolas Hiot \orcidlink{0000-0003-4318-4906}\\
LIFO -- Universit\'e d'Orl\'eans, INSA CVL -- Orl\'eans, France\\
EnnovLabs -- Ennov -- Paris, France
\\~\\
Dominique Laurent \orcidlink{0000-0002-7264-9576}\\
ETIS -- CNRS, ENSEA, CY Universit\'e -- Cergy-Pontoise, France}

\maketitle

\begin{abstract}
Efficient consistency maintenance of incomplete and dynamic real-life databases is a quality label for further data analysis. 
In prior work, we tackled the generic problem of database updating in the presence of tuple generating constraints from a theoretical viewpoint. 
The current paper considers the usability of our approach by  (a) introducing incremental update routines (instead of the previous from-scratch versions) and (b) removing the restriction that limits the contents of the database to fit in the main memory.
In doing so, this paper offers new algorithms, proposes queries and data models inviting discussions on the representation of incompleteness on databases.
We also propose implementations under a graph database model and the traditional relational database model. 
Our experiments show that computation times are similar globally but point to discrepancies in some steps.
\end{abstract}

\section{Introduction}
	\label{secintro}
	
Incremental update algorithms are essential for incomplete real-life databases,  often large and constantly updated.
Modern applications usually involve the analysis of large amounts of data with missing and changing values.
The quality of this analysis depends on the consistency of the data, the maintenance of which requires calculations whose cost needs to be reduced.

To address this problem, we build upon our prior work~\cite{CHL20}
where the generic problem of database updating in the presence of constraints was tackled from a theoretical point of view, and under the restriction that the database content was meant to fit in main memory.
Hence, for missing values, we follow  Reiter~\cite{Rei86} who provides FOL (First-Order Logic) semantics to null values of type `value exists but is currently unknown'. 
Constraints  are expressed as
tuple-generating dependencies (tgd), \ie  implications of the form $(\forall X,Y)(B(X,Y) \Rightarrow (\exists Z)L(X,Z))$ where $X$, $Y$, $Z$ are vectors of variables, $B(X,Y)$ is the conjunction of atoms of the form $L_i(X_i, Y_i)$ where $X_i$ and $Y_i$ are sub-vectors of $X$ and $Y$, respectively, and $L(X,Z)$ is an atom. 

In~\cite{CHL20}, our purpose was to allow for the insertions or the deletions of sets of tuples under the following hypotheses:

\begin{itemize}
	\item A fixed set $\Ct$ of constraints  as specified just above  is assumed over a given set of predicates.
	\item The database $\Db$ to be updated is a set of instantiated atoms, in which marked nulls may occur. Moreover, the database contains no redundancies caused by these nulls.
	\item $\Db$ satisfies the constraints in $\Ct$, meaning that, for every $c$ in $\Ct$, whenever there exists an instantiation $h$ of $X$ and $Y$ such that $\Db$ contains all atoms in $h(B(X,Y))$ then $h$ can be extended to $Z$ so as $\Db$ also contains  $h(L(X,Z))$.
	\item The updated database $\Db'$ satisfies the constraints in $\Ct$ and is not  redundant, while being such that the updates are performed, that is, all atoms to be inserted are present in $\Db'$ and no atoms to be deleted is present in $\Db'$.
\end{itemize}

%
In this paper, we improve our work in~\cite{CHL20} in two main aspects:
(1) we  propose an \textit{incremental} version of the approach and
(2) we deal with data stored in \textit{database systems},  contrary to the in-memory version of~\cite{CHL20}.

Given  an update $u$ over a database instance \Db, 
our approach consists in generating (by the activation of constraints in \Ct) a set of new updates, $U'$, as necessary side-effects  to maintain the database consistent. Contrary to  \textit{from scratch} algorithms,  whereby \textit{the whole } database instance and \textit{the whole} set \Ct\ are  involved in the generation of $U'$,  \textit{incremental} algorithms minimize the amount of data to be accessed and the constraints to be triggered.
 
 \vspace{0.25cm}
 \noindent
 \textsc{Paper Organisation.}
 We overview our approach and its evolution (\textit{from scratch} towards an \textit{incremental} approach)
  through a motivating example in Section~\ref{sec:run-ex}. 
Section~\ref{sec:back} provides some background.
Section~\ref{sec:simplification} introduces the operations over which the incremental core computation is built.
Incremental update algorithms,  their implementation aspects and 
experimental results are introduced, respectively in
Sections~\ref{sec:IncUp}, \ref{sec:Impl} and ~\ref{sec:expe}.
After presenting related work in Section~\ref{sec:RW}, Section~\ref{sec:conclusion} concludes the paper.

\section{Motivating Example}
	\label{sec:run-ex}
	Figure~\ref{fig:runex-constraints} shows a set of constraints in the context of a university, researchers and students. Although the intuitive meaning of these constraints should be clear, we point out that in the constraints $c_6$, $c_9$, $c_{10}$, $c_{11}$ and $c_{12}$, the right hand-side involves a variable not present in the left hand-side. Due to such contraints, known in the literature as tuple generating dependencies (tgd), nulls values may appear in the database instance, as explained below. Their intuitive meaning is as follows:
\begin{itemize}
	\item the variable $Z$ in $PhDPaper(Y,P,Z)$ of $c_6$ stands for the year the paper has been published;
	\item the variable $Y$ in $Cites(X,Y)$ of $c_9$ stands for a publication cited  by publication $X$;
	\item the variable $Y$ in $Enrolled(X,Y)$ of $c_{10}$ stands for a course student $X$ is enrolled in;
	\item the variable $Z$ in $Degree(Y,Z)$ of $c_{11}$ stands for the degree course $Y$ is part of;
	\item the variable $Z$ in $Language(X,Y,Z)$ of $c_{12}$ stands for the language in which is taught course $X$ of degree $Y$.
\end{itemize}

\begin{figure}[t] 
    \centering
    \small{
	\fbox{\parbox{.97\linewidth}{
	\begin{multicols}{2}
	\begin{enumerate}[label=$c_{\arabic*}:$,ref=$c_{\arabic*}$]
		\item \label{c1} $Supervises(X, Y) \rightarrow Researcher(X)$
		\item \label{c2} $Supervises(X, Y) \rightarrow Student(Y)$
		\item \label{c3} $Authors(X, Y) \rightarrow Researcher(X)$
		\item \label{c4} $Authors(X, Y) \rightarrow Publication(Y)$
		\item \label{c5} $Researcher(X) \rightarrow Authors(X, Y)$
		\item \label{c6} $Supervises(X, Y), Authors(X, P),$ \\ $Authors(Y, P) \rightarrow PhDPaper(Y, P, Z)$
		\item \label{c7} $Cites(X, Y) \rightarrow Publication(X)$
		\item \label{c8} $Cites(X, Y) \rightarrow Publication(Y)$
		\item \label{c9} $Publication(X) \rightarrow Cites(X,Y)$
		\item \label{c10} $Student(X) \rightarrow Enrolled(X, Y)$
		\item \label{c11} $Enrolled(X, Y) \rightarrow Degree (Y, Z)$
		\item \label{c12} $Degree (X, Y) \rightarrow Language (X,Y,Z)$
		\item \label{c13} $Enrolled(X, Y) \rightarrow GrantEligible(X)$
	\end{enumerate}
	\end{multicols}}
	}}
	\caption{Set of (general) constraints}  \label{fig:runex-constraints}
\end{figure}

Constraints from this set are used in subsequent examples to illustrate our proposes throughout the paper. Let us start with  $\Ct=\{c_1, \ldots ,c_6\}$ over the following database instance $\Db$:

\smallskip\noindent
\begin{tabular}{rl}
$\Db = \{$&$Researcher(Elin), Authors(Elin, P_{269}), Publication(P_{235}), Authors(Sten, P_{269}),$\\
	& $Publication(P_{269}), Student(Sten), Supervises(Elin, Sten), Researcher(Nils),$ \\
	& $PhDPaper(Sten, P_{269}, 2022)~\}$
\end{tabular}

\vspace{.2cm}\noindent
{\sc (A) Constraint satisfaction.~}
First,  constraints $c_1$ and $c_2$ are satisfied, because $Supervises(Elin,$ $Sten)$, $Researcher(Elin)$ and $Sudent(Sten)$ are in $\Db$. Constraint $c_4$ is satisfied as well because  $Publication(P_{269})$ is  in $\Db$.
Similarly, $c_6$ is  satisfied   because $\Db$ contains $PhDPaper(Sten,$ $P_{269},2022)$.

However, $c_3$ is {\em not} satisfied because $Authors(Sten, P_{269})$ is in $\Db$ and  $Researcher(Sten)$ is not. Similarly, $c_5$ is not satisfied because $Researcher(Nils)$ has no matching fact over $Authors$ in $\Db$.
Constraint satisfaction is obtained by adding facts, generated by a process  called \textit{chase}:
\begin{enumerate}
	\item $Researcher(Sten)$  is added to satisfy  $c_3$.
	\item In order to satisfy $c_5$ one fact over $Authors$ must be added, but  the value of the second argument (\ie \,the publication) is unknown. Despite that publications are present in $\Db$, those authored by $Nils$ are unknown. {\em Marked nulls} are used to account for this situation: $Authors(Nils, N_1)$ is added, which is read as `$Nils$ authored a publication, currently unknown but recognized as $N_1$'. The atom $Publication(N_1)$ is then inserted in order to satisfy $c_4$.
\end{enumerate}
In this case, we obtain $\Db' = \Db \cup \{Researcher(Sten),$ $Authors(Nils, N_1),$ $Publication(N_1))\}$.

In the following items, we show differences between the \textit{from-scratch} and incremental approaches to updates.

\vspace{0.2cm}\noindent
{\sc (B) Updates.~}
Updates are insertions or deletions.

\smallskip\noindent
{\sc (B.1)} {\bf Insertions.} Given a database instance $\Db$ satisfying a set of constraints $\Ct$, an updated database is the result of inserting facts in $\Db$ while maintaining constraint satisfaction.

Let $\Ct_1 = \{c_1, c_6, c_7,c_8\}$, $\Db_1=\{Researcher(Elin), $ $Super\-vises(Elin, Sten), $ $Authors(Elin, P_{269})\}$, and the set of required insertions $\ins =\{Authors(Sten, P_{269})\}$.

\smallskip\noindent
\underline{\textit{From-scratch} approach.} To reinforce constraints, side-effects are computed through a process called $chase$, that applies the constraints in $\Ct_1$ on $\Db_1 \cup \ins$, to generate a new set $chase(\Db_1)$. In doing so, facts already in $\Db_1$ might be generated again.

\smallskip\noindent
\underline{Incremental approach.} Side-effects  for insertions are computed based on  $\ins$ as follows. In our example,
the only constraint to be triggered when inserting $Authors(Sten, P_{269})$ is  $c_6$ because (i) a query on $\Db_1$
informs that the atoms $Supervises(Elin, Sten)$ and $Authors(Elin, P_{269})$ already exist in $\Db_1$, and (ii) $c_6$ is impacted by the insertion.
Then, the result  of the insertion is $\Db_1' = \Db_1 \cup \{Authors(Sten, P_{269}),$ $PhDPaper(Sten, P_{269}, N_2)\}$.

\vspace{.2cm}\noindent
{\sc (B.2) }{\bf Deletions.}
Consider $\Db'$  (from item (A) above)  along with the constraints in $\Ct=\{c_1, \ldots , c_6\}$. 

\smallskip\noindent
\underline{\textit{From-scratch} approach.} 
After removing from $\Db'$ a given set of facts $\del$,  the deletion process includes the computation of 
$chase (\Db' \setminus \del)$ to check constraint satisfaction. Then, if one atom to be deleted is re-generated (up to null renaming),
a backward chase is activated to identify the side-effects of the
deletion. This is illustrated through the following two cases.

\smallskip\noindent
{\em Case 1.} Let $\del = \{PhDPaper(Sten,P_{269}, 2022)\}$.
First the fact $PhDPaper(Sten,P_{269}, 2022)$ is removed from $\Db'$ and then, constraint satisfaction is checked as done for insertions.  Here, $\Db' \setminus\{PhDPaper(Sten,P_{269}, 2022)\}$ does not satisfy $c_6$ because Sten is still present as  an author of paper $P_{269}$.
  As above, consistency is restored by inserting $PhDPaper(Sten,P_{269}, N_2)$. The resulting database  $\Db''= (\Db' \setminus$ $\{PhDPaper(Sten,P_{269}, 2022)\})~ \cup$ 
$\{PhDPaper(Sten,P_{269}, N_2)\}$
is consistent and implements the deletion because it does not contain the atom to be deleted.

\smallskip\noindent
{\em Case 2.} Consider now $\del = \{PhDPaper(Sten,P_{269}, N_2)\}$ on $\Db''$.
A processing similar to the previous one would first remove the atom from $\Db''$ and then, 
insert $ PhDPaper(Sten,P_{269},  N_3)$ to restore consistency with respect to the constraints.
This result is not acceptable because the generated set is equal to $\Db''$ (up to a null renaming), meaning that the deletion has {\em not} been implemented. 
 In this case, the processing is carried on by deleting all atoms responsible of the generation of the atom to be deleted.
 
 This amounts to apply  the constraints {\em backward} (from the head to the body), removing \textbf{one} atom from the body, to prevent the constraint from being triggered.
To this end,  for every constraint $c$, a literal in its body is marked as the one to be deleted in such a situation
(for the sake of simplicity, let  the leftmost literal in be the marked one).
Here, $Authors(Elin, P_{269})$ has to be deleted, due to $c_{6}$. 
Then, we proceed following the ideas already presented:
the latter deletion, because of $c_{5}$, requires the insertion of $ Authors(Elin, N_4)$ which in turn, 
because of $c_4$, requires the insertion $ Publication(N_4)$, returning the database

\smallskip
$\Db_2=(\Db'' \setminus\{PhDPaper(Sten,P_{269}, N_2), Authors(Elin, P_{269})\})~\cup$\\
\rightline{$\{Authors(Elin, N_4), Publication(N_4)\}$.}

\smallskip\noindent
\underline{Incremental approach.} First, a backward chase is used to find the constraints impacted by the required update.
In our example, constraint $c_{6},$ and then $ c_{5}$ are the only ones concerned by deletions.
Second, the chase is applied \textit{only} on the rules detected just above and its result is analyzed as done in the \textit{from-scratch} approach.
It is worth noting that,  as data is stored in a  database,  queries are used to detect whether constraints can be applied for chasing (backward or forward).

\vspace{0.2cm}\noindent
{\sc (C) Avoiding too many nulls.~}
An important issue regarding side effects is making sure that the processing terminates. Considering the set of  constraints $\Ct_1=\Ct \cup \{c_7,c_8,c_9\}$ on $\Db'$, it is clear that $\Db''$ does {\em not} satisfy $c_9$, because $Publication(P_{269})$ and $Publication(P_{235})$ belong to $\Db''$ with no associated citation.
In order to satisfy $c_9$, $Cites(P_{269}, N_6)$ and $Cites(P_{235}, N_7)$ must be inserted, which triggers the insertions of $Publication(N_6)$ and of $Publication(N_7)$, due to $c_7$. Then, to satisfy $c_9$, $Cites(N_6, N_8)$ and $Cites(N_7,N_9)$ have to  be inserted, and  we are clearly entering an infinite loop, which is not acceptable.

To cope with this difficulty, every null $N$ is associated with an integer called {\em the degree of $N$} and denoted by $\delta(N)$. At each insertion, the degree of all nulls occurring in $\Db$ are set to $0$, and when a constraint $c$ is applied during the processing, all generated nulls are assigned a degree equal to $\delta +1$ where $\delta$ is the maximal degree of the nulls in the atoms of the constraint body, or $0$ if no null occurs in the constraint body. Moreover, assuming a fixed maximal null degree $\delta_{\max}$, insertion processing is {\em stopped} as soon as a null $N$ is such that $\delta(N) \geq \delta_{\max}$, and the insertion is {\em rejected}, that is $\Db$ is not changed.

For example, in the case just above, we have $\delta(N_6)=\delta(N_7)=0$, $\delta(N_8) = \delta(N_6)+1=1$ and $\delta(N_9) = \delta(N_7)+1=1$, etc. If, for example $\delta_{\max}$ is set to $2$, the generation of nulls will stop at the next round and the insertion will be \textbf{rejected}.  
The verification of null degree  is similar in both {\em From-scratch} and incremental approaches (it was proven in \cite{CHL20} that by using $\delta_{max}$ we accept only  consistent insertions).

\vspace{0.2cm}\noindent
{\sc (D) Avoiding redundancies (core).~}
Side effects have to be computed in a {\em minimal} way to reflect as much as possible the so-called {\em minimal change} requirement.
To illustrate this point consider the insertion in $\Db'$ of $Authors(Nils, P_{235})$. Adding this fact in $\Db'$ provokes redundancies, because  the presence of $Authors(Nils, N_1)$ and of $Publication(N_1)$ is no longer required to ensure constraint satisfaction. The result of this insertion is the set $\Db'''$ defined by:

\smallskip
	$\Db'''= (\Db' \setminus \{Authors(Nils, N_1), Publication(N_1)\}) \cup \{Authors(Nils, P_{235})\}$

\smallskip\noindent
In our implementations, redundancies in $\Db$ are eliminated through the computation of the {\em core}, seeking for  mapping nulls to constants or nulls so as to detect redundant atoms. In our example, for $h$ such that $h(N_1)=P_{235}$, we have:\\
$-$ $h(Authors(Nils, N_1))=Authors(Nils, P_{235})$ and\\
$-$ $h(Publication(N_1))=Publication(P_{235})$,\\
showing that $Authors(Nils, N_1)$ and $Publication(N_1)$ are redundant.

\smallskip\noindent
\underline{\textit{From-scratch} approach.} Once the updates are performed on the database, the whole instance is considered for simplifications.

\smallskip\noindent
\underline{Incremental approach.} This new proposal aims to retrieve  only the facts involved in the update operation. 
For instance, for $\Db'$ as in our  example, in the incremental approach a query detects that $N_1$ is  the \textit{only} null value concerned by the update.
No need to work with the whole instance $\Db'$.

\paragraph{From-scratch and incremental approaches at a glance.}
Consider the update process that  includes  the general ideas explained in items (B) and (C) above.
Denote, respectively, by \textit{upd} and $upd_{|U}$,  its \textit{from-scratch} and \textit{incremental} versions.
More precisely, when using  the  $upd_{|U}$ policy,  only  the database portion impacted by  $U$ is concerned, while the whole database is concerned by $upd$ policy.
The expression $\Db \diamondsuit U$ indicates the insertion/deletion of the required updates $U$ in/from $\Db$.

In the \textit{from-scratch} approach the new instance is denoted by $\Db' = core(upd(\Db \diamondsuit U))$, 
while in  the incremental approach, the new instance is denoted by $\Db' = core_{|NullBucket}(upd_{|U}(\Db\diamondsuit U))$, where \textit{NullBucket}, is the set of nulls impacted by the update policy ($upd_{|U}$) applied to $\Db \diamondsuit U$.

\section{Preliminaries}
\label{sec:back}
We recall some formal definitions already used in~\cite{CHL20}.
We assume a standard FOL alphabet composed of three pairwise disjoint sets, namely: {\sc const}, a set of constants, {\sc var}, a set of variables and {\sc pred}, a set of predicates, every predicate being associated with a positive integer called its arity. 
In this setting, a {\em term} is a constant or a variable and an atomic formula, or an atom, is a formula of the form $P(t_1, \ldots , t_n)$ where $P$ is a predicate of arity $n$ and $t_1, \ldots , t_n$ are terms. Every atom in which no variables occur is called a {\em fact}.

A {\em homomorphism} from a set of atoms $A_1$ to a set of atoms $A_2$ is a mapping $h$ from the terms of $A_1$ to  the  terms of $A_2$ such that:
$(i)$ if $t\in $ {\sc const}, then $h(t)= t$, and  $(ii)$ if $P(t_1,...,t_n)$ is  in $A_1$, then $P(h(t_1), . . . , h(t_n))$ is in $A_2$. 
The set $A_1$ is {\em isomorphic} to the set $A_2$ if there exists a homomorphism $h_1$ from $ A_1$ to $A_2$ which admits an inverse homomorphism (from $A_2$ to $A_1$).

We denote by $\Phi$ the set of all formulas of the form $(\exists \X)(\varphi_1(\X_1) \wedge \ldots \wedge \varphi_n(\X_n))$ where $\X$ is a vector of variables made of all variables occurring in $\X_i$ ($i=1, \ldots , n$), and where for every $i=1, \ldots ,n$, $\varphi_i(\X_i)$ is an atomic formula in which the free variables are those in $\X_i$. If $\phi$ denotes such a formula in $\Phi$, the set $\{\varphi_1(\X_1), \ldots , \varphi_n(\X_n)\}$ is denoted by $atoms(\phi)$.

Given $\phi$ in $\Phi$, a {\em model} $M$ of $\phi$ is a set of facts such that there exists a homomorphism from $atoms(\phi)$ to $M$. In such a setting,  for all $\phi_1$ and $\phi_2$ in $\Phi$, $\phi_1 \Rightarrow \phi_2$ holds if each model of $\phi_1$ is a model of $\phi_2$, and as usual, $\phi_1$ and $\phi_2$ in $\Phi$ are said to be {\em equivalent}, denoted by $\phi_1 \Leftrightarrow \phi_2$, if $\phi_1 \Rightarrow \phi_2$ and $\phi_2 \Rightarrow \phi_1$ both hold, that is if $\phi_1$ and $\phi_2$ have the same models.

For all $\phi_1$ and $\phi_2$ in $\Phi$, $\phi_1$ is said to be {\em simpler than} $\phi_2$, denoted by $\phi_1 \preceq \phi_2$, if
$(i)$ $\phi_1 \Leftrightarrow \phi_2$ holds, and $(ii)$ $atoms(\phi_1) \subseteq atoms(\phi_2)$.
$\phi_1$ is also said to be a {\em simplification} of $\phi_2$. A simplification $\phi_1$ of $\phi_2$ is said to be {\em minimal} if $\phi_1 \preceq \phi_2$ and  there is no $\phi'_1$ such that $\phi'_1 \prec \phi_1$.
For instance, let $\phi$ be the formula $(\exists x,y)(P(a,x) \wedge P(a,y))$; then $(\exists x)(P(a,x))$ and $(\exists y)(P(a,y))$ are two distinct but {\em equivalent} simplifications of $\phi$.

It is shown in~\cite{CHL20} that 
if $\phi$ is in $\Phi$ and $\phi_1$ and $\phi_2$ two {\em minimal} simplifications of $\phi$, then $atoms(\phi_1)$ and $atoms(\phi_2)$ are isomorphic
(in the literature, we find a similar result for graphs\cite{HellN92}). 
Minimal simplifications are also called {\em cores} and the core of a given formula $\phi$ is denoted by $core(\phi)$. 

Basically, a \textit{database instance} is a formula $\phi$ in $\Phi$ that cannot be simplified, \ie such that $core(\phi)=\phi$. 
Formulas in $\Phi$ are  `skolemized' by replacing the variables with specific constants referred to as {\em Skolem constants} or as {\em (marked) nulls} and by omitting the existential quantifier.
We thus assume an additional set of symbols in our alphabet, denoted by {\sc null},  disjoint from the sets {\sc const} and {\sc var}.
Now a term can be of one of the following types: either a constant, or a null, or a variable. 
Any atom of the form $P(t_1, \ldots ,t_n)$ where for every $i=1 , \ldots ,n$, $t_i$ is in ${\mbox{\sc const}} \cup {\mbox{\sc null}}$, is called an {\em instantiated atom}.
Given an instantiated  atom $A$, denote by $null(A)$  the set of nulls  appearing in $A$.
Moreover, as usual, the transformed conjunctive formula is written as the {\em set} of its conjuncts. In other words, \textit{a database instance is a set of instantiated atoms that can be written as $atoms(Sk(\phi))$ where $Sk(\phi)$ is the Skolem version of a formula $\phi$ in $\Phi$ such that $core(\phi)=\phi$.}

\section{Simplification with Respect to Nulls: a Basic Operation}
\label{sec:simplification}

In our approach, a database $\Db$ is expected to be equal to its core to avoid data redundancy.
 It is thus of paramount importance to enforce this property when updating. To this end, we propose {\em incremental} algorithms, so as to deal with as few nulls  as possible, based on those  involved in the update processing.
 
More formally, given a set of atoms $I$  and a set of nulls $\nu$ occurring in $I$, we look for a homomorphism $h$ 
such that for every $N$ not in $\nu$, $h(N)=N$ and $h(I)$ is minimal so as  $h(I) \subseteq I$. However, the following example shows that the choice of $\nu$ cannot be arbitrary. Indeed, given a set of nulls $\nu_0$, with respect to which $I$ is to be simplified, the set $\nu_0$ has to be expanded to the set $\nu$ of all nulls `linked' (directly or indirectly) to a null in $\nu_0$ in some atom of $I$.
\begin{example}
\label{ex:simplify1}
{\rm
Let  $\nu_0= \{N_1\}$ and $I$ defined by:

\smallskip\noindent
\begin{tabular}{rl}
$I=\{$&$Student(Alice), Enrolled(Alice, N_1), Degree (N_1, N_2), Enrolled(Alice, Math),$\\
	&$Degree (Math,N_3), Degree(CS, N_4), Degree(CS, BSc)~\}$
\end{tabular}

\smallskip\noindent
To simplify $I$ \wrt\ $\nu_0$,  we should eliminate  redundancies in  $I$ involving $N_1$. As $N_1$ occurs in $Degree (N_1, N_2)$ with the other null $N_2$, the simplification should deal with $N_1$ {\em and} $N_2$. Since $N_1$ and $N_2$ are not linked with any other null in the atoms of $I$, we have $\nu=\{N_1,N_2\}$. For $h$ such that $h_1(N_1) =  Math$ and $h_1(N_2)= N_3$, we obtain a non-redundant instance $I'=h(I)$ defined by

\smallskip\noindent
\begin{tabular}{rl}
$I'=\{$&$Student(Alice), Enrolled(Alice, Math),Degree (Math, N_3),$\\&$ Degree(CS, N_4), Degree(CS, BSc)~\}.$
\end{tabular}

\smallskip\noindent
Notice however that simplifications involving $N_3$ or $N_4$ have {\em not} to be considered.
\hfill$\Box$}
\end{example}
As shown by the above example, given $I$ and $\nu_0$, nulls `linked' in $I$ to nulls in $\nu_0$ have to identified. We do so through the computation for every $N$ in $\nu_0$, of the set ${\sf LinkedNull}_{I,N}$ as explained next.
%
We first define the sequence $\left( \textsf{LinkedNull}^k_{I,N} \right)_{k \geq 0}$ by:
\begin{enumerate}[label={(\roman*)}]
	\item $\textsf{LinkedNull}^0_{I,N} = \{ A_i \in I \mid N \in null(A_i)\}$
	\item $\textsf{LinkedNull}^k_{I,N} = \{ A_i \in I \mid (\exists A_j \in \textsf{LinkedNull}^{k-1}_{I,N}) (null(A_i) \cap null(A_j) \neq \emptyset)\}$.
\end{enumerate}
It is easy to see that for  every $k \geq 0$, we have
$\textsf{LinkedNull}^k_{I,N} \subseteq \textsf{LinkedNull}^{k+1}_{I,N}$ and 
$\textsf{LinkedNull}^k_{I,N} \subseteq I$.
Thus, the sequence $\left( \textsf{LinkedNull}^k_{I,N} \right)_{k \geq 0}$
is bounded by $I$ and is monotonic. As $I$ is finite, the sequence has a unique limit, which is precisely the sub-set  of $I$ denoted by
 $\textsf{LinkedNull}_{I,N}$.
 
 It therefore turns out that redundancy has only to be checked with respect to the atoms in $\bigcup_{N \in \nu_0} \textsf{LinkedNull}_{I,N}$ and the set $\nu$ of all nulls occurring in this set. Algorithm~\ref{coreMaintenance-DL} shows how redundancies are dealt with in this context.

{\small
	\algsetup{indent=1.5em}
	\begin{algorithm}[th]
		\caption{$Simplify (I, \nu_0)$\label{coreMaintenance-DL}}
		\begin{algorithmic}[1]
			\STATE{$PSet := \{\textsf{LinkedNull}_{I,N} \mid N \in \nu_0\}$}\label{core:findLinkedNulls-DL}
			\FORALL{\label{core:loopLinkNulls-DL} $P \in PSet$}
				\STATE{Build the query $q_{core}$ and compute its answer $q_{core} (I)$}\label{corequery-DL}
				\IF{\label{core:multAns-DL} $\mid  (q_{core} (I))  \mid > 1$}
					\STATE {$h_m:=$ \textit{ChooseMostSpec}($q_{core}(I)$)}\label{coreMostSpec-DL}
						\STATE{$I:= (I \setminus P) \cup h_m(P)$} \label{core:delP-DL}  
				\ENDIF 
			\ENDFOR
		\RETURN $I$
		\end{algorithmic}
	\end{algorithm}
}

Algorithm~\ref{coreMaintenance-DL}  receives  as input a set $I$ of instantiated atoms, and a set of nulls $\nu_0$.
For each $N$ in $\nu_0$, the algorithm computes the set $\textsf{LinkedNull}_{I, N}$ 
 (line~\ref{core:findLinkedNulls-DL}), which is stored in a set called $PSet$. Therefore, the nulls occurring in $PSet$ constitute the set $\nu$ with respect to which $I$ is simplified.
 
On line~\ref{corequery-DL}, for each $P$ in $PSet$, a query $q_{core}: ans (X) \leftarrow A_1(X_1), \dots, A_n (X_n)$ is built by replacing each occurrence of $N_i$ in $P$ by $x_i$. That is, $A_i(X_i)$ is obtained from $A_i$ in $P$ by replacing the nulls in $A_i$ by the corresponding variables.
 
Thus, assuming that $p$ nulls occur in $P$, when evaluating the answer $q_{core}(I)$  of $q_{core}$, the tuple $(N_1, \ldots , N_p)$ is obviously returned. However, it may happen that the answer contains other tuples, each of which define a possible instantiation of the nulls in $P$. In this case, some atoms in $P$ are redundant, and thus can be removed.
To implement these remarks, when the evaluation of  $q_{core}$ over $I$ returns more than one tuple (line~\ref{core:multAns-DL}),
one most specific tuple is chosen  (line~\ref{coreMostSpec-DL}), and denoting by $h_m$ the associated homomorphism, $I$ is simplified (line~\ref{core:delP-DL}) by replacing all atoms $A$ in $P$ by $h_m(A)$.
\begin{example}
\label{ex:simplifyFindqcore}
{\rm
Considering $I$ as in Example~\ref{ex:simplify1} and $\nu_0= \{N_1\}$, $ \textsf{LinkedNull}_{I, N_1} $ consists of the atoms
$Enrolled(Alice, N_1)$ and $Degree(N_1, N_2)$.
Thus, the query  $q_{core}$  is defined by:

\smallskip
	$ans(x_1, x_2) \leftarrow Enrolled(Alice, x_1), Degree(x_1, x_2)$

\smallskip\noindent
returning the answer $\{(N_1, N_2), (Math, N_3)\}$ with more than one tuple. Hence, $h_m$ such that $h_m(N_1) = Math$ and $h_m(N_2)= N_3$ is returned line~\ref{coreMostSpec-DL}, and $I$ is simplified as illustrated in Example~\ref{ex:simplify1}. \hfill$\Box$}
\end{example}
%
To explain our method for computing the most specific homomorphism $h_m$ we introduce the notion of  $P$-homomorphism.
\begin{definition}
Given $I$ a set of instantiated atoms and  $N$ a null occurring in $I$, let $P=\textsf{LinkedNull}_{I,N}$. A $P$-homomorphism is a homomorphism $h$  such that $h(I) \subseteq I$ and for every null $N'$ in $null(I) \setminus null(P)$, $h(N')=N'$.

$I$ is said to be $P$-reducible if there exists a $P$-homomorphism $h$ such that $h(I)$ is a strict subset of $I$. $\hfill~\Box$
\end{definition}


\noindent
In the following proposition, given a set $I$  of instantiated atoms and a null $N$ in $\nu_0$, we use the following notation:
\begin{itemize}
\item $P$ denotes the set of atoms $\textsf{LinkedNull}_{I,N}$, and $null(P)=\{N_1, \ldots N_p\}$ denotes the set of nulls occurring in $P$;
\item $q_{core}(I)$ is the answer to $q_{core}$ computed against $I$. That is, $q_{core}(I)$ is the set $\{h_1, \ldots, h_q\}$ of all possible $P$-homomor\-phisms defined over $null(P)$. We suppose that $h_1$ is the identity, {\ie ~} for every $j=1,\ldots ,p$, $h_1(N_j)=N_j$;
\item $H_P$ denotes the table with $p$ columns and $q$ rows such that $H_P[i,j] = h_i(N_j)$.
\item Given a set of atoms $Q$, we denote by $cons\_null(Q)$ the set of all symbols $\sigma$ such that $\sigma$ is a constant or a null {\em not} in $null(Q)$.
\end{itemize}
We  recall that given two homomorphisms $h_1$ and $h_2$ over the same set of symbols $\Sigma$, $h_1$ is said to be {\em less specific than} $h_2$, denoted by $h_1 \preceq h_2$, if there exists a homomorphism $h$ over $\Sigma$ such that $h \circ h_{1}=h_2$. Using these notation, the following proposition holds.
\begin{proposition}\label{prop:spec}
Given $h_i$ and $h_{i'}$ in $q_{core}(I)$,  $h_i \preceq h_{i'}$ holds if and only if, for every $j=1 , \dots , p$, we have:
\begin{enumerate}
\item
If $H_P[i,j]$ is in $cons\_null(P)$, then $H_P[i,j] = H_P[i',j]$;
\item
If $H_P[i,j]$ is a null $N$ in $null(P)$, then for every $j' \ne j$ such that $H_P[i,j] = H_P[i,j']$ it holds that $H_P[i',j] = H_P[i',j']$.
\end{enumerate}
\end{proposition}
{\sc Proof.}
Let us first assume that $h_i \preceq h_{i'}$ holds. In this case, there exists $h$ such that $h \circ h_{i} = h_{i'}$. If $N_j$ is such that $h_i(N_j)$ is a constant or a null not  in $null(P)$, then for every $P$-homomorphism $h_P$, $h_P(h_{i}(N_j))= h_{i}(N_j)$. Hence, $h_{i'}(N_j)= h \circ h_{i}(N_j) = h_{i}(N_j)$, which shows item (1). If $j$ and $j'$ are such that $h_i(N_j)= h_i(N_{j'})$, then $h_{i'}(N_j)= h_{i'}(N_{j'})$ also holds, showing item (2).

Conversely, assume  that for $h_i$ and $h_{i'}$, items (1) and (2) hold. Let $h$ be defined for every $j = 1, \ldots , p$ as follows: if there exists $N_k$ such that $h_i(N_k)=N_j$ then $h(N_j) = h_{i'}(N_k)$, otherwise $h(N_j)=N_j$.  We first notice that $h$ is well defined. Indeed, if $k$ and $k'$ are such that $h_i(N_k)=h_i(N_{k'})$, then we have two expressions defining $h(N_j)$, namely $h(N_j) = h_{i'}(N_k)$ and $h(N_j) = h_{i'}(N_{k'})$. However, by item (2) we have $h_{i'}(N_k)=h_{i'}(N_{k'})$, and thus, these two expressions yield the same value. We now prove that $h_{i} \preceq h_{i'}$, that is, that for every   $k = 1, \ldots , p$,  then $h_{i'}(N_k)=h(h_{i}(N_k))$. If $h_i(N_k)$ is not in $null(P)$, then, we have $h_i(N_k)=h_{i'}(N_k)=N_k$, and by construction of $h$ we also have $h(N_K)=N_k$. Therefore $h_{i'}(N_k)=h(h_{i}(N_k))=N_k$. On the other hand, if $h_i(N_k)=N_j$, by definition of $h$, we have $h_{i'}(N_k)=h(N_j)$. Hence, $h_{i'}(N_k)=h(N_j)=h(h_i(N_k))$. Since for every  $k = 1, \ldots , p$, we have $h_{i'}(N_k)=h(N_j)=h(h_i(N_k))$, it follows that $h_i \preceq h_{i'}$, and the proof is complete.
\hfill$\Box$
%
%
\begin{example}\label{ex:most-spec}
{\rm
Let $I=\{B(N_1,N_2),$ $B(N_2,N_1),$ $C(N_1,a),$ $C(N_2,a),$ $C(N_3,a)\}$ and $\nu_0=\{N_1\}$.

In this case, $P=\{\{B(N_1,N_2),$ $B(N_2,N_1),$ $C(N_1,a),$ $C(N_2,a)\}\}$ and thus $null(P)=\{N_1, N_2\}$ and $cons\_null(P)=\{a, N_3\}$.
This implies that $P$-homomorphisms  should {\em not} change $N_3$, or in other words, $N_3$ should be treated as constant. The query $q_{core}$ is thus written as follows:

$q_{core}: ans(x_1,x_2) \leftarrow B(x_1,x_2),B(x_2,x_1),C(x_1,a),C(x_2,a)$

\smallskip\noindent
and the table $H_P$ representing the answer $q_{core}(I)$ is shown below.

\begin{center}
\begin{tabular}{c|cc}
$H_P$& $x_1$ & $x_2$\\ \hline  
1 & $N_1 $& $N_2$\\ 
2 & $N_2 $& $N_1$\\ 
\end{tabular}
\end{center}
$H_P$ has 2 columns (because $null(P)$ contains two nulls), and 2 rows due to two answers in $q_{core}(I)$.
It is easy to see that  $h_1 \preceq h_2$, and $h_2 \preceq h_1$ 
meaning that there is no advantage in trying to simplify the database instance in this case.
Indeed,  we have $h_1(I) = I$,  where $h_1$ is the identity. 
We have $h_2(I) = I$ as well, although $h_2 $ is not the identity.
Remark that $h_2$ does not satisfy $h_2 = h_2 \circ h_2$ (\ie $h_2$ is not idempotent) because  $h_2(h_2(N_1))=h_2(N_2)=N_1$, whereas $h_2(N_1)=N_2$. As will be seen shortly, detecting such homomorphisms allows for computational optimizations.
\hfill$\Box$}
\end{example}
The following corollary shows how to find one most specific homomorphism, based on Proposition~\ref{prop:spec}. To state the corollary, we use the following notation for  $i=1, \ldots ,q$:
\begin{itemize}
\item
$\gamma_i$ is the number of nulls $N$ in $null(P)$ such that $h_i(N)$ is in $cons\_null(P)$;
\item
$\mu_i=\{k \in \{1, \ldots, q\}~|~(\forall j=1, \ldots, p) (h_i(N_j) \in cons\_null(P) \Rightarrow h_k(N_j) = h_i(N_j))\}$;
\item
$\pi_i$ is the number of distinct nulls in $null(P)$ in the set $h_i(null(P))$.
\end{itemize}
Intuitively speaking, considering that $H_P$  is the tableau,   then $\gamma_i$ is the number of columns that, at row $i$, contain a symbol in $cons\_null(P)$. On the other hand, $\mu_i$ is the set of all rows in $H_P$ containing the same symbols of $cons\_null(P)$ in the same columns as row $i$ does (\ie\, if $h_i(N_j) = c$ then $h_k(N_j)=c$). Then $\pi_i$ is the number of distinct nulls in $nulls(P)$ occurring in row $i$.

The corollary below formalizes the following informal remarks:
\begin{enumerate}
	\item If $h_i \ne h_i \circ h_i$, then $h_i$ cannot be one of the most specific homomorphisms, because in this case, $h_i \prec h_i \circ h_i$. For instance, in Example~\ref{ex:most-spec}, we have  $h_2 \neq h_2 \circ h_2$.

	\item Most specific homomorphisms  are among the rows of $H_P$ with the largest number of symbols in $cons\_null(P)$. 
Indeed, let $h_i$ and $h_j$ be such that  row $i$ contains strictly more symbols in $cons\_null(P)$ than row $j$ and $h_i \prec h_j$. Then, there exists $h$ such that $h \circ h_i = h_j$, and so, if $N$ in $null(P)$ is such that $h_i(N)$ is in $cons\_null(P)$, we have $h(h_i(N))=h_i(N)$, and so $h_i(N)=h_j(N)$. 
Thus,  row $j$ has at least as many symbols in $cons\_null(P)$ as row $i$, which implies a contradiction. Hence, for every $N$ in $null(P)$, $h_i(N)$ is also in $null(P)$, in which case rows $i$ and $j$ have no symbols in $cons\_null(P)$, which is another contradiction.
%

	\item Considering one of the rows defined just above, say row $i$, among all rows having the same symbols in  $cons\_null(P)$ in the same columns as row $i$, we argue that a row with as few distinct nulls in $null(P)$ defines one most specific homomorphism.
\end{enumerate}

\begin{corollary}\label{coro: spec}
Given $I$ and $P$ as above, denoting by $\{h_1, \ldots, h_q\}$ the set $q_{core}(I)$, the following holds:
\begin{enumerate}
\item
If $h_i$ is one of the most specific $P$-homomorphisms in $q_{core}(I)$ then $h_i$ is idempotent, that is, $h_i \circ h_i = h_i$. 
\item
$h_{i}$ is one of the most specific $P$-homomorphisms in $q_{core}(I)$ if (a)  $\gamma_i = \max_{1 \leq j\leq q}(\gamma_j)$, and
(b) $\pi_i = \min_{k\in \mu_i}(\pi_k)$.
\end{enumerate}
\end{corollary}
{\sc Proof.}
First, Proposition~\ref{prop:spec} implies that $h_i \preceq h_i \circ h_i$ holds for every $h_i$. Moreover, as $h_i$ is a $P$-homomorphism, we have  $h_i(I) \subseteq I$. Thus $h_i \circ h_i(I) \subseteq I$, which implies that $h_i \circ h_i$ is a $P$-homomorphism as well. The proof of  item (1) is therefore complete.

Assume that $h_i$ satisfies (2) and let $h_k$ be a $P$-homomorphism such that $h_i \preceq h_k$. By Proposition~\ref{prop:spec}, if $h_i(N_j)$ is in $cons\_null(P)$, then $h_i(N_j)=h_k(N_j)$. Therefore, $\gamma_i \leq \gamma_k$, and as $\gamma_k \leq \gamma_i$, this implies $\gamma_i = \gamma_k$. Thus, $k$ is in $\mu_i$, which implies that $h_i(N_j)$ is in $null(P)$ if and only if so is $h_k(N_j)$. By Proposition~\ref{prop:spec}, if $j$ and $j'$ are such that $h_i(N_j)=h_i(N_{j'})$ then we also have $h_k(N_j)=h_k(N_{j'})$. It therefore follows that less nulls in $null(P)$ occur for $h_k$, that is $\pi_k \leq \pi_i$. As $\pi_i \leq \pi_k$ must hold, we obtain that $\pi_i = \pi_k$, meaning that $h_i$ and $h_k$ are equal up to a null renaming. The proof is therefore complete.
\hfill$\Box$
{\small
	\algsetup{indent=1.5em}
	\begin{algorithm}[th]
		\caption{$ChooseMostSpecific(q_{core}(I))$\label{mostspecific}}
		\begin{algorithmic}[1]
		
			\STATE{Build $H_P$ as explained in Proposition~\ref{prop:spec}}\\
			\COMMENT{$H_P$ has $q$ rows and $p$ columns}
			\STATE{$row\_max:=1$ ; $count\_max := 0$ ; $i:=2$}
			\FORALL{$i=2, \ldots , q$}\label{line:loop1-spec}
				\STATE{$idemPot := {\tt true}$}
				\STATE{$count\_curr := 0$ ; $j := 1$}
				\WHILE{$idemPot = {\tt true}$ and $j\leq p$}
					\IF{$H_P[i,j]$ is in $cons\_null(P)$}
						\STATE{$count\_curr := count\_curr +1$}
					\ELSE
						\STATE{Let $N_k = H_P[i,j]$ \COMMENT{$N_k$ is in $null(P)$}}
						\IF{$H_P[i,k] \ne N_k$}
							\STATE{$idemPot := {\tt false}$ \COMMENT{$h_i$ is not idem-potent}}
							\STATE{Mark row $H_P[i]$}\label{line:non-idem}
						\ENDIF
					\ENDIF
					\STATE{$j:=j+1$}
				\ENDWHILE
				\IF{$idemPot = {\tt true}$}
					\IF{$count\_curr > count\_max$}\label{test:count-max}
						\STATE{$row\_max:=i$}\label{assing:row-max}
						\STATE{$count\_max := count\_curr$}
					\ENDIF
				\ENDIF
			\ENDFOR
			\STATE{$row\_spec:=row\_max$}\label{assign:row-spec1}
			\STATE{Let $count\_min$ be the number of distinct nulls in $null(P)$ occurring in $H_P[row\_max]$}\label{assing:row-min}
			\FORALL{$i=2, \ldots , q$}	\label{line:loop2-spec}
				\IF{row $H_P[i]$ is not marked and $i \ne row\_max$}
					\STATE{$match := {\tt true}$ ; $j:=1$}
					\WHILE{$match = {\tt true}$ and $j\leq p$}
						\IF{$H_P[row\_max,j]$ is in $cons\_null(P)$ and $H_P[row\_max,j] \ne H_P[i,j]$}
							\STATE{$match := {\tt false}$}
						\ENDIF
						\STATE{$j:=j+1$}
					\ENDWHILE
					\IF{$match = {\tt true}$}\label{line:test-match}
						\STATE{Let $count\_null$ be the number of distinct nulls in $null(P)$ occurring in $H_P[i]$}
						\IF{$count\_null < count\_min$}
							\STATE{$row\_spec := i$}\label{assign:row-spec}
							\STATE{$count\_min := count\_null$}
						\ENDIF
					\ENDIF
				\ENDIF
			\ENDFOR
			\RETURN{$h_m$, the homomorphism defined by $H_P[row\_spec]$}
		\end{algorithmic}
	\end{algorithm}
	\normalsize
}

\noindent
As a consequence, finding a most specific $P$-homomorphism in $q_{core}(I)$ amounts to $(i)$ discard any row not defining an idem-potent homomorphism and $(ii)$ among the remaining rows, identify one homomorphism satisfying item 2 of Corollary~\ref{coro: spec}. Algorithm~\ref{mostspecific} shows how to compute such a most specific homomorphism, and we notice that this does {\em not} require data  access. To end the section, we illustrate Algorithm~\ref{mostspecific} as follows.
%
\begin{example}\label{ex:most-spec-DL}
{\rm
We first consider the context of Example~\ref{ex:most-spec}, where $I=\{B(N_1,N_2),$ $B(N_2,N_1),$ $C(N_1,a),$ $C(N_2,a),$ $C(N_3,a)\}$ and 
$P=\{B(N_1,N_2),$ $B(N_2,N_1),$ $C(N_1,a),$ $C(N_2,a)\}$.

In this case, $null(P)=\{N_1, N_2\}$, $cons\_null(P)=\{N_3\}$, and the associated table $H_P$ has been shown already.
Applying Algorithm~\ref{mostspecific} based on the table $H_P$, the following computations are achieved.

The first loop line~\ref{line:loop1-spec} aims at marking rows defining non idempotent $P$-homomorphisms (that is, such that $h \circ h \ne h$) and mean-while to find one unmarked row with as many symbols in $cons\_null(P)$ as possible, in reference to Corollary~\ref{coro: spec}(2). These computations return the following:
\begin{itemize}
\item
When processing row $2$ of $H_P$, we have $H_P[2,1]=N_2$ where $N_2$ is in $null(P)$, and  $H_P[i,2]=N_1$. Since $N_1 \ne N_2$, $idemPot$ is set to ${\tt false}$ and row $2$ is marked on line~\ref{line:non-idem}.
\item
Since there is no other row to process, the loop line~\ref{line:loop1-spec} returns $row\_max = 1$ and $count\_curr =0$.
\end{itemize}
Hence, Algorithm~\ref{mostspecific} returns $row\_spec=1$ and so,  $h_m$ is defined by $h_m(N_1)=N_1$ and $h_m(N_2)=N_2$. In other words, $I$ is not simplified, which is indeed the expected result.

\smallskip
We now illustrate further Algorithm~\ref{mostspecific}, using two more sophisticated cases.
First, let $\nu_0=\{N_1\}$ and $I_1=\{B(N_1, N_2),$ $B(a, N_2),$ $B(a,N_3),$ $B(N_4,N_3),$ $C(N_2,N_2),$ $C(N_3,N_3)\}$.
In this case, $\textsf{LinkedNull}_{I,N_1}=\{N_1,N_2\}$ and thus, $PSet = \{P\}$ where $P=\{B(N_1, N_2),$ $B(a,N_2),$ $C(N_2,N_2)\}$. Moreover,  the query

\smallskip
$q_{core}:ans(x_1,x_2) \leftarrow B(x_1, x_2), B(a,x_2),C(x_2,x_2)$

\smallskip\noindent
is generated and its answer against $I_1$, $q_{core}(I_1)$, is defined in the following table $H^1_P$:

\begin{center}
\begin{tabular}{c|cc}
$H_P^1$& $x_1$ & $x_2$\\ \hline  
1 & $N_1 $& $N_2$\\ 
2 & $a$& $N_2$\\ 
3 & $a$&$ N_3$\\ 
4 & $N_4$ & $ N_3$\\ 
\end{tabular}
\end{center}
$H_P^1$ has 2 columns and 4 rows due to four possible answers in $q_{core}(I_1)$. Moreover,
$h_1 \preceq h_2$,  $h_2 \preceq h_3$ and $h_2 \preceq h_4$. Notice that
$h_3$ and $h_4$ are not comparable because $a, N_3$ and $N_4$ are in $cons\_null(P)$.
Applying Algorithm~\ref{mostspecific} based on the table $H_P^1$, the first loop line~\ref{line:loop1-spec} achieves the following:
\begin{itemize}
\item
No row is marked as non-idempotent on line~\ref{line:non-idem}. This is so because for every $i=1, \ldots , 4$, and every $j=1,2$, if $H^1_P[i,j]=N_k$ where $N_k$ is $N_1$ or $N_2$, $H^1_P[i,k]=N_k$.
\item Regarding the value of $count\_curr$, the computed value is $0$ for the first row, $1$ for row $2$, and $2$ for rows $3$ and $4$ (because $a, N_3$ and $N_4$ are in $cons\_null(P)$). Thus, applying the test line~\ref{test:count-max}, $count\_curr$ to set to $2$, and on line~\ref{assing:row-max}, $row\_max$ is set to $3$. Indeed, although for row $4$, we have  $count\_curr = 2$,  the test line~\ref{test:count-max} fails, and thus the value of $row\_max$ is not changed. Then, $row\_spec$ is set to $3$ on line~\ref{assign:row-spec1} and $cont\_min$ is set to $0$ on line~\ref{assing:row-min}.
\end{itemize}
Therefore, processing the loop line~\ref{line:loop2-spec} yields no change and Algorithm~\ref{mostspecific} returns $h_m$ defined by $h_m(N_1)=a$ and $h_m(N_2)=N_3$, in which  case,  $h_m(I_1)=\{B(a, N_3), B(N_4,N_3), C(N_3,N_3)\}$, which is not redundant, when considering $N_3$ and $N_4$ as particular `constants'.

\smallskip
As a more sophisticated illustration of Algorithm~\ref{mostspecific}, let $\nu_0=\{N_1\}$ and $I_2= \{B(N_1, N_2),$ $B(a, N_2),$ $C(N_2,N_2),$ $C(N_2,N_3)\}$.
Here, $\textsf{LinkedNull}_{I,N_1}=\{N_1,N_2, N_3\}$ and thus, $PSet = \{P\}$ where $P=\{B(N_1, N_2),$ $B(a,N_2),$ $C(N_2,N_2),$ $C(N_2,N_3)\}$. 
Moreover,  the query:

\smallskip
$q_{core}:ans(x_1,x_2,x_3) \leftarrow B(x_1, x_2), B(a,x_2),C(x_2,x_2),C(x_2,x_3)$

\smallskip\noindent
is generated and $q_{core}(I_2)$, is defined in the following table $H^2_P$:

\begin{center}
\begin{tabular}{c|ccc}
$H_P^2$& $x_1$ & $x_2$ & $x_3$\\ \hline  
1 & $N_1 $& $N_2$ & $N_3$ \\ 
2 & $N_1$& $N_2$ & $N_2$\\ 
3 & $a$& $N_2$ & $N_3$ \\ 
4 & $a$& $N_2$ & $N_2$ \\
\end{tabular}
\end{center}
$H_P^2$ has 3 columns and 4 rows due to four possible answers in $q_{core}(I_2)$. Moreover,
$h_1 \prec h_2$,  $h_1 \prec h_3$, $h_2 \prec h_4$ and $h_3 \prec h_4$.
Applying Algorithm~\ref{mostspecific} based on the table $H_P^2$, the loop line~\ref{line:loop1-spec} achieves the following:
\begin{itemize}
\item
As above, no row is marked as non idempotent on line~\ref{line:non-idem}. This is so because for  $i=1, \ldots , 4$, and  $j=1,2,3$, if $H^2_P[i,j]=N_k$ where $N_k$ is $N_1$, $N_2$ or $N_3$, $H^2_P[i,k]=N_k$.
\item
As above, on line~\ref{assing:row-max}, $row\_max$ is set to $3$ and thus, $row\_spec$ is set to $3$ on line~\ref{assign:row-spec1}. Here, $count\_min$ is set to $2$ on line~\ref{assing:row-min} because $N_2$ and $N_3$ are in $null(P)$.
\end{itemize}
When processing the loop line~\ref{line:loop2-spec}, the only row to be considered is row $4$, for which $match$ is {\tt true}, thus implying that the test on line~\ref{line:test-match} succeeds. Since for row $4$, the value of $count\_null$ is $1$ (because row $4$ contains the only null $N_2$), the value of $row\_spec$ is set to $4$, line~\ref{assign:row-spec}. Hence, Algorithm~\ref{mostspecific} returns $h_m$ defined by $h_m(N_1)=a$, $h_m(N_2)=N_2$ and $h_m(N_3)=N_2$.
In this case,   $h_m(I_2)=\{B(a, N_2), C(N_2,N_2)\}$, which is not redundant.
\hfill$\Box$}
\end{example}

Homomorphisms have been used in database theory during the last decades, in the field of query optimization \cite{ASU79,CM77} (we refer to  \cite{AHV95} for an overview). We notice in this respect that, in \cite{ASU79}, a partial pre-ordering between homomorphisms is defined using the same criteria as in Proposition~\ref{prop:spec}, showing that our approach to simplification is closely related to the field of query optimization. 
Roughly, in our approach, we  compare  all the answers ($h_1, \dots h_q$)  for  $q_{core}$ and chose one ($h_m$) among the most  specific ones
(which  are incomparable).
From another point of view, if we consider queries  $Q_1$, $Q_2$, $\dots$, $Q_q$ as the instantiations of $q_{core}$ by  $h_1, \dots h_q$, respectively, then $h_m$ is a  homomorphism such that  $h_m(body (Q_i)) = body (Q)$  for all $Q \subseteq Q_i$.
Actually, our simplification technique is based on tableau optimization, as done in \cite{ASU79} for query optimization, where  the sets of variables and of distinguished variables are, respectively, called, in our approach, the $null(P)$ and $cons\_null$. However, the contexts and the expectations in our approach are fundamentally different from those  summarized in~\cite{AHV95}. 
Indeed: 
\begin{itemize}
\item
In \cite{AHV95}, the tableau is built up from the query {\em body}, whereas in our approach, the tableau is built up from the answer to a given query.
\item
Our approach generates {\em one} most specific homomorphism, where as the approach shown in \cite{AHV95} aims at discarding all non most specific.
\end{itemize}
As a result, the problem we deal with can be seen as {\em more specific} than the general case considered in \cite{AHV95, ASU79}, thus resulting in a specific algorithm. 

\section{Incremental Updating}
\label{sec:IncUp}

In~\cite{CHL20}, update algorithms work on in-memory data, using no  DataBase Management System (DBMS).
This version considers  a DBMS, based on which data access is implemented through queries. 
In this section, we show how to implement updates  by restricting  data access as much as possible. 

\begin{figure}
\begin{center}
\begin{tabular}{l|c|l}
Query & Algo & Purpose \\ \hline
$q_{bucket}(I)_{[S]}$ &\ref{insertionAlgo}, \ref{delAlgo}&  retrieves all nulls in $I$ appearing in an atom\\
				& &$p(...)$ such that $p$ is a predicate in a given set $S$\\
$q_{degree}(I)_{[S,\delta_{max}]}$	&\ref{insertionAlgo}& for each $N$ in $S$,  checks if  $N$ is in $I$ and if\\
&&$\delta(N) <  \delta_{max}$  \\
$q_{\delta}(I)_{[S,d]}$& \ref{insertionAlgo}& for each $N$ in $S$,  if  $N$ is in $I$, sets $\delta(N)$ to $d$\\
$q_{Iso}(I)_{[S]}$ & \ref{delAlgo} & retrieves in $I$ all atoms isomorphic to those in $S$ \\

\end{tabular}
\noindent
\caption{\label{fig:queries}
Queries used in our algorithms}
\end{center}
\end{figure}

\subsection{Insertion}
\label{secIns}
Algorithm~\ref{insertionAlgo} describes the insertion in $\Db$ of the atoms in the set $\ins$. 
On line~\ref{ins:chase}, the side-effects of the insertion  are computed and stored in the set \textit{ToIns}, and then the instance $\Db'=\Db \cup ToIns$ is simplified on line~\ref{ins:core} through the computation of its core. If all nulls in the simplified instance have a degree less than  the specified maximal degree $\delta_{max}$ (on line~\ref{ins:testNull}), null degrees are all set to $0$ (on line~\ref{ins:setsNull0}) and $\Db'$ is returned since, as shown in~\cite{CHL20}, it is always consistent; otherwise, the database is not modified.



{\small
	\algsetup{indent=1.5em}
	
	\begin{algorithm}[!ht]
		\caption{Insert($\Db , \Ct, \delta_{max}, \ins)$ \label{insertionAlgo}}
		\begin{algorithmic}[1]

			\STATE \label{ins:chase} $ToIns := Chase4Insert(\Db , \Ct, \delta_{max}, \ins)$
			\STATE  \label{ins:bucket} $NullBucket := \{N_j \mid N_j$ is a null obtained by $q_{bucket}(\Db \cup ToIns)_{[ToIns]} \}$

			\STATE \label{ins:core} $\Db' := Simplify(\Db \cup ToIns, NullBucket)$

			\IF {\label{ins:testNull} $q_{degree}(\Db')_{[NullBucket,\delta_{max}]}$ }
			\STATE{\label{ins:setsNull0} $q_{\delta}(\Db')_{[NullBucket,0]}$}
			\RETURN  $\Db'$ 
			\ELSE
			\RETURN $\Db$
			\ENDIF

		\end{algorithmic}
	\end{algorithm}
}

Contrary to the algorithms  in~\cite{CHL20}, the main steps in Algorithm~\ref{insertionAlgo} are designed in an {\em incremental} manner.
%
First, to avoid generating any non necessary side effect atoms,
an {\em incremental} version of the chase procedure considered. According to this procedure, a constraint $c$ is activated \textit{only} when the following conditions hold:

\smallskip
(i) $body(c)$ contains at least one  atom that maps to one being inserted, and

(ii)  atoms in $body(c)$ that do not respect  (i) map to  atoms in the database instance \Db.

\smallskip\noindent
This new chase differs from the one in~\cite{CHL20} in the following aspects:
(a) only the rules $c$  concerned by insertions  are triggered 
and  (b) queries are built to find  in \Db\ instantiations for atoms in $body(c)$.

{\small
	\algsetup{indent=1.5em}
	\begin{algorithm}[t]
		\caption{Chase4Insert($ \Db, \Ct,  \delta_{max}, \ins )$\label{chaseAlgoIns}}
		\begin{algorithmic}[1]

			\STATE {\label{chA:initSet}  $ToIns := \ins$}
			\WHILE {\label{chA:loop} $\exists c \in \Ct $  and  $\exists h$ such that 
			$h(body(c)) \subseteq \Db \cup ToIns$ and $h(body (c)) \cap ToIns \neq \emptyset$ and 
			$\delta (h'(head(c)) \leq \delta_{max}$, where $h' \supseteq h$ maps to new nulls all existential variables in $head(c)$, \\
			and there does not exist $h'' $ such that $ h''(h'(head(c)) \in \Db \cup ToIns$}
			\STATE {\label{chA:finalSet}$ToIns := ToIns \cup \{h'(head(c))\}$\\
			\COMMENT{Degrees of new nulls in $h'(head(c))$ are set to  $d_{\max} +1$, where $d_{\max}$ is the maximal\\degree  in $h(body(c))$, or $0$  if $h(body(c))$ contains no null}}
						\ENDWHILE
			\RETURN $ToIns$
			
		\end{algorithmic}
	\end{algorithm}
}

Algorithm~\ref{chaseAlgoIns}, called on line~\ref{ins:chase}, implements our incremental chase procedure.
The set $ToIns$ initially stores   $\ins$ (line~\ref{chA:initSet})  and then, stores the generated  side-effects (line~\ref{chA:finalSet})
through the while loop on line~\ref{chA:loop},  defined by the following conditions:
\begin{itemize}
\item A constraint $c$ is triggered \textit{only} if  at least one atom in $body(c)$ is instantiated to an atom in $ToIns$.
\item The condition $\delta(h'(head(c)) \leq \delta_{max}$ ensures that only (side-effect) atoms whose degree is less than the maximum null degree are kept.
The instantiation $h'$ extends $h$ by assigning new null values to existential variables in $head(c)$. When performing a chase step, the degree of new nulls are also computed.
\item The last condition ensures termination along with a simplification.
Indeed, if an instantiation of $h'(head(c))$, referred to as $h''(h'(head(c)))$, exists in $\Db \cup ToIns$, the constraint is satisfied, and no atom is inserted in $ToIns$.
For instance, suppose $\Db_1 = \{Authors(Elin,P_2)\},$ $\Ct = \{c_5\}$ (Figure~\ref{fig:runex-constraints}) and
$\ins = \{Researcher(Elin)\}$. 
The atom $Authors (Elin, N_1)$, generated by $c_5$, is not inserted since it maps to $Authors(Elin,P_2)$.
\end{itemize}
Another difference between the algorithm in \cite{CHL20} and Algorithm~\ref{insertionAlgo}, is the simplification step on line~\ref{ins:core}  to maintain the database instance irredundant. Indeed, based on our earlier discussion in Section~\ref{sec:simplification},  $\Db \cup ToIns$ is simplified
 \wrt\  the nulls  in \textit{NullBucket}, computed through the query  $q_{bucket}$ on line~\ref{ins:bucket}. Thus, only the nulls in \textit{NullBucket} and their `linked' nulls are considered, thus optimizing the computation of the core of  $\Db \cup ToIns$.
  \begin{example}
 \label{ex:Bucket-Core-Null}
 {\rm
 Let $\Ct = \{c_1, c_3, c_4,c_{10}, c_{11}, c_{12}\}$, $\delta_{max} = 3$ and
 the following database instance:

 \smallskip\noindent 
 \begin{tabular}{rl}
  $\Db = \{$ &$Authors(N_1, P_2),Authors(Alice, N_2),Publication(P_2), Publication(N_2), $\\
 & $Researcher(N_1),Researcher(Alice), Supervises (N_1, N_3)~\}$
   \end{tabular}
 
 \smallskip\noindent 
 Let $\ins= \{Authors(Alice,  P_5), $ $Stu\-dent(Bob)\}$. Running Algorithm~\ref{insertionAlgo} in this case is as follows.
Constraint $c_4$ is triggered due to the insertion of $Authors(Alice,$ $ P_5)$
 and constraints $c_{10}, c_{11}, c_{12}$ are triggered due to the insertion of $Student(Bob)$.
 Line~\ref{ins:chase} returns the following set $ToIns$, where null degrees are shown as exponents:
 
 \smallskip\noindent 
\begin{tabular}{rl}
 $ToIns = \{$&$Authors(Alice,P_5), Publication(P_5), Student(Bob), Enrolled (Bob, N_5^0), $\\
 &$Degree(N_4^0,N_5^1), Language(N_4^0, N_5^1, N_6^2)~\}$.
 \end{tabular}

\smallskip\noindent
To simplify $\Db \cup ToIns$, the query $q_{Bucket}$ 
retrieves in  $\Db$ the nulls concerning $Authors$ (\ie $N_1$ and $N_2$),
$Publication$ (\ie $N_2$), $Enrolled$ (\ie $N_4$), $Degree$ (\ie $N_4$, $N_5$) and $Language$ (\ie $N_4$, $N_5$, $N_6$).
Therefore, $NullBucket =\{N_1,$ $N_2,$ $N_4,$ $N_5,$ $N_6\}$, and by Algorithm~\ref{coreMaintenance-DL}, we obtain that
$\textsf{LinkedNulls}_{\Db, N_1}= \{N_1, N_3\}$,
$\textsf{LinkedNulls}_{\Db, N_2}=\{N_2\}$, and for $i=4,5,6$, $\textsf{LinkedNulls}_{\Db, N_i}=\{N_4,N_5,N_6\}$.
The simplification of $\Db \cup ToIns$
(line~\ref{ins:core} of Algorithm~\ref{insertionAlgo})
results in:

\smallskip\noindent
\begin{tabular}{rl}
 $\Db' =\{$&$Authors(N_1, P_2), Authors(Alice, P_5), Publication(P_2), Publication(P_5), Researcher(N_1),$\\
&$Researcher(Alice), Supervises(N_1,N_3), Student(Bob), Enrolled (Bob, N_4),$\\
&$Degree(N_4,N_5),Language(N_4, N_5, N_6)~\}$.
\end{tabular}

\smallskip\noindent
Notice also that, since the degree of nulls is checked only during the chase, before returning the updated instance,
the degrees of all nulls are set to $0$ on line~\ref{ins:setsNull0} of Algorithm~\ref{insertionAlgo}.\hfill$\Box$
}
\end{example}

\subsection{Deletion}
\label{secDel}

{\small
	\algsetup{indent=1.5em}
	
	\begin{algorithm}[th]
		\caption{$Delete( \Db, \Ct, \delta_{max}, \del)$  \label{delAlgo}}
		\begin{algorithmic}[1]
	
			\STATE \label{del:iso} $\delIso := q_{Iso}(\Db)_{[\del]}$
			\COMMENT{$\delIso$ contains  atoms  
			in $\Db$  that have to be deleted}
			\STATE \label{del:chase} $ToDel, ToIns := Chase4Delete ( \Db, \Ct,\delta_{max}, \delIso )$
			\STATE {\label{del:buildInstance} $\Db' := (\Db \cup ToIns) \setminus ToDel$}
		
			\STATE{ \label{del:bucket} $NullBucket := \{N_j \mid N_j$ is a null obtained by $q_{bucket}(\Db')_{[ToIns \cup ToDel]}\}$}
			\STATE \label{del:core} $\Db' := Simplify(\Db', NullBucket)$
			\RETURN $\Db'$
		\end{algorithmic}
	\end{algorithm}
}

Our incremental algorithm for the deletion from $\Db$ of atoms in $\del$ is displayed as Algorithm~\ref{delAlgo}. On line~\ref{del:iso}, all atoms in $\Db$ isomorphic  to one in the set $\del$ are retrieved through the query $q_{iso}$. 
For instance, if  $\del = \{P(a, N_1)\}$ and $\Db_1= \{P(a, N_5)\}$, then query $q_{iso}$ returns $\{P(a, N_5)\}$.
The side-effects are then computed on line~\ref{chaseAlgoDel},  recalling from Section~\ref{sec:run-ex} that the side effects involve not only atoms to be deleted, but also atoms to be {\em inserted} as side-effects. In Algorithm~\ref{delAlgo}, the corresponding sets are respectively denoted by $ToDel$ and $ToIns$.

Once these side-effects have been incorporated in $\Db$ to produce $\Db'$ (line~\ref{del:buildInstance}), this new instance is simplified as in the case of insertion: impacted nulls are generated on line~\ref{del:bucket} and the simplified instance is computed on line~\ref{del:core}. We notice that, contrary to insertions, deletions are {\em never} rejected.

As for insertions, side effects are computed {\em incrementally} through  Algorithm~\ref{chaseAlgoDel}. First, it may happen that
%
the deletion of an instantiated atom $A$  makes the database inconsistent when it is a consequence of a constraint $c$.
To find all such constraints $c$, we reason backward on \Ct\  to find an instantiation $h$ such that $h(head(c))=A$.
Then $h$  is extended to verify, in a forward reasoning,  whether $body(c)$ can be triggered and generate $A$ again.

The idea  is to check whether $c$ generates an atom isomorphic to an atom being deleted (Algorithm~\ref{chaseAlgoDel}, line~\ref{chasedel:iso1}).
If so, at least one atom in $h(body(c))$ should be deleted in order to prevent $c$ from being triggered. This atom is then inserted in $ToDel$ (line~\ref{chasedel:toDel}). Notice that, to avoid  non-determinism, it is assumed that the atom to be deleted has been marked as `$-$' during rule design.

If no atom isomorphic to an atom to be deleted is generated,  a new set called $NewToIns$ is generated as the side-effects of inserting the new instance of $head(c)$ and  all atoms in $ToIns$ (line~\ref{chasedel:chIns}). If no atom in $NewToIns$ meets an atom to be deleted and if the degrees of the involved nulls are less that $\delta_{max}$, then these atoms are inserted in $ToIns$ (line~\ref{chasedel:endIf2}). Otherwise, the marked atom from $h(body(c))$ is inserted in $ToDel$ (line~\ref{chasedel:else2}).

{
	\small
	\algsetup{indent=1.5em}
	\begin{algorithm}[th]
		\caption{Chase4Delete($\Db, \Ct,  \delta_{max}, \delIso)$\label{chaseAlgoDel}}
		\begin{algorithmic}[1]
			\STATE $ToIns := \emptyset$ and $ToDel := \delIso$
			\WHILE{\label{chasedel:while}there exist $c \in \Ct$ and $h$ such that  $h(head(c)) \in ToDel$ and $h(body(c)) \subset (\Db \setminus ToDel) \cup ToIns$}
				\IF{\label{chasedel:iso1}$\exists h'$ such that $h'(body(c)) = h(body(c))$ and $h'(head(c))$ is isomorphic\\to $h(head(c))$}
				    \STATE{\label{chasedel:toDel}$ToDel := ToDel \cup \{h'(body^-(c))\}$}
			     \ELSE \label{chasedel:beginElse1}
			     \STATE{\label{chasedel:chIns} $ NewToIns := \textsf{Chase4Insert}(\Db, \Ct, \delta_{max}, ToIns \cup \{h'(head(c))\})$}
			       \IF  {\label{chasedel:If2} $NewToIns=Del = \emptyset$ and  $\delta(N)< \delta_{max}$ for all nulls $N$ in $NewToIns$  }
					\STATE{\label{chasedel:endIf2} $ToIns = ToIns \cup NewToIns$}
				  \ELSE
					\STATE {\label{chasedel:else2}$ToDel := ToDel \cup \{h'(body^-(c))\}$}
				\ENDIF
				\ENDIF
			\ENDWHILE
			\RETURN $ToDel, ToIns$
		\end{algorithmic}
	\end{algorithm}
}

\begin{example}
{\rm
Let  $\Db_0=\{GrantEligible(Sten), $ $Student(Sten),$ $Enrolled(Sten,CS)\}$,   $\Ct = \{c_{10},$ $c_{13}\}$ and
$\del = \{Grant\-Eligible(Sten)\}$.

\smallskip
On line~\ref{del:chase}, Algorithm~\ref{delAlgo} calls  Algorithm~\ref{chaseAlgoDel} to perform an incremental chase. 
$ToIns$ and $ToDel$ are respectively initialized to $\emptyset$  and $\{GrantEligible(Sten)\}$, and a first iteration of the loop on line~\ref{chasedel:while} is run.
Constraint $c_{13}$ is concerned by the deletion, because for $h$ such that
$h(head(c_{13}))= GrantEligible(Sten)$,  as $Enrolled(Sten, CS)$ is in $\Db$,  $c_{13}$ generates $Grant\-Eligible(Sten)$ (line~\ref{chasedel:iso1}).
Therefore,  $ToDel$ is set to $\{GrantEligible(Sten),$ $Enrolled$ $(Sten,CS)\}$ and $ToIns$ remains empty.

In the second iteration of the loop, $c_{10}$ is detected to be concerned by the deletion of the atom $Enrolled(Sten,CS)$.
With $Student(Sten)$ in $\Db$,  $c_{10}$ generates $Enrolled(Sten, N_1)$, which is not isomorphic to  $Enrolled(Sten,CS)$ (line~\ref{chasedel:iso1}).
The next step consists in testing  whether the atom $Enrolled(Sten, N_1)$ should be added to $ToIns$.
To this end, Algorithm~\ref{chaseAlgoDel} chases forward, starting with $Enrolled(Sten, N_1)$ (line~\ref{chasedel:chIns}) to generate 
 $GrantEligible(Sten)$. This atom being in $ToDel$ (line~\ref{chasedel:If2}), $Student(Sten)$ is added to $ToDel$, and nothing is added in $ToIns$.
 Algorithm~\ref{chaseAlgoDel} returns $ToDel = \{Student(Sten), Grant\-Eligible(Sten), Enrolled(Stem,CS), Enrolled(Sten,N_1)\}$, and $ToIns = \emptyset$.
 Algorithm~\ref{delAlgo} then performs the deletions and the resulting database instance is empty. $\hfill\Box$
  
}
\end{example}

\section{Queries for Incremental Processing}
\label{sec:Impl}
By implementing our method using graph and relational database models, our goal is to study performance aspects, and to raise 
issues concerning the database design regarding queries.
\subsection{Graph Data Model}
\label{sec:graph}
The DBMS considered in this work is  Neo4J, which deals with attributed graphs. 
Cypher is a well-established language for querying and updating \textit{property graph databases}. As explained in~\cite{FGGLLMPRS18}, `a Cypher query takes as input a property graph and outputs a table. These tables can be thought of as providing bindings for parameters that witness some patterns in a graph, with some additional processing done on them'.
The central concept in Cypher queries is pattern matching. 
The MATCH clause searches for homomorphisms identifying a given pattern in the queried graph. 
The returned result is an instance over a  table where attributes correspond to the variables  in the Cypher query.

Our approach involves managing null values that have to be retrieved based on their co-occurrences as arguments of atoms (Section~\ref{sec:simplification}). 
Given a null $N_1$ we need to efficiently detect atoms having $N_1$ as one of its arguments, and then for every $N$ occurring with $N_1$, to recursively access the atoms having $N$ as argument. In doing so, the set of nulls is partitioned into blocks whose elements are those nulls that have to be considered in the simplification steps.
To make such retrieval efficient, 
we adopt a model close to the logical formalism used  in our previous explanations, composed of three types of nodes. 
Given an atom $P(t_1, \dots, t_n)$ 
our graph database represents $P$ as a node, linked to other nodes representing the terms
$t_1, \dots, t_n$. 
In this context, nodes in our graph database are of three possible types distinguished by \textit{labels}, and all nodes have \textit{properties}, among which one is \textbf{symbol}. More precisely:
\begin{itemize}
	\item Nodes of type \textit{Atom} have one label \textbf{:Atom} representing the predicate symbol in an atom. This predicate symbol is the value of property \textbf{symbol} of such a node.
	\item Nodes of type \textit{Constant} representing constant values. Such nodes have two labels, \textbf{:Element} and \textbf{:Constant},  and the value of their property \textit{symbol} is the constant itself.
	\item Nodes of type \textit{Null} reprensenting nulls. Such nodes have two labels, \textbf{:Element} and \textbf{:Null}, and the value of their property \textbf{symbol} is the name of the null prefixed with `$\_$'.
\end{itemize}
%
An edge links nodes with label \textbf{:Atom} to nodes with label \textbf{:Element}.
Moreover, an edge has the property \textit{rank},  allowing to refer  to the terms of an atom by their positions.

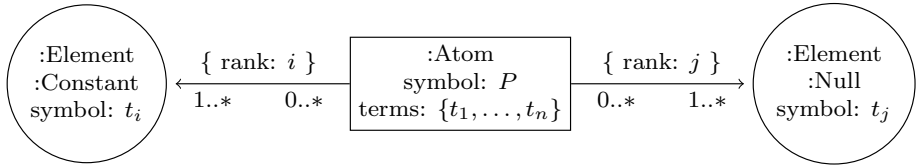
\begin{figure}[htb]
	\centering
	\resizebox{.8\linewidth}{!}{
	\footnotesize
		\begin{tikzpicture}[shorten >=3pt,->,node distance=4.5cm,on grid,every text node part/.style={align=center}]
			\node[draw] (a) {:Atom\\symbol: $P$\\terms: $\{t_1, \dots, t_n\}$}; 
			\node[draw,circle,left = of a] (nc) {:Element\\:Constant\\symbol: $t_i$}; 
			\node[draw,circle,right = of a] (nn) {:Element\\:Null\\symbol: $t_j$}; 
			\path
			(a) edge [] node[below, near start] {$0..*$} node[below, near end] {$1..*$} node[above] {\{ rank: $i$ \}} (nc)
			(a) edge [] node[below, near start] {$0..*$} node[below, near end] {$1..*$} node[above] { \{ rank: $j$ \}} (nn)
			;
		\end{tikzpicture}
	}
	\caption{Graph database schema. \label{fig:schema-graph}}
\end{figure}

Figure~\ref{fig:schema-graph}  illustrates the schema  of the atom $P(t_1, \dots, t_n)$ by representing
 constant terms by $t_i $ and nulls by $t_j$. 
 Notation below edges indicates the cardinality of  the relationship between an atom and its terms:
 an element is connected to at least one atom and  atoms may have no terms.

Figure~\ref{fig:runex-graph} illustrates part of our database instance (rectangular nodes are atoms and circular nodes are elements).

As explained before, our model benefits certain operations.
However, it increases the cost of conversions between the graph-format and the logic-format for an atom.
Such conversions are essential for the communication between the database and  the  procedures  performing some computations locally.
To optimize these conversions and graph traversals, we introduce the following redundancies in our database model,
which  have significantly improved our implementation (see Section~\ref{sec:expe}).

\begin{itemize}
\item \textit{To avoid edge traversal.}
\begin{enumerate}[label=(\alph*)]
   \item[] For each node \textbf{:Atom}, we store, as  its attribute,  an ordered list containing all its terms. 
In  Figure~\ref{fig:schema-graph}, the rectangular node shows this new attribute: \textit{terms}.
For example, to obtain atom $Authors(Elin, $ $P_{269})$ from the instance in Figure~\ref{fig:runex-graph} starting with the node
$n_{117}$, instead of traversing edges $r_{19}$ and $r_{20}$, we just have to retrieve the attributes \textit{terms} of node $n_{117}$.
%
 
\end{enumerate}

\item \textit{To allow efficient access to nodes.}
\begin{enumerate}[label=(\alph*)]
	\item A uniqueness constraint is added on the \textbf{Element} \textit{symbol} (implying, \eg, that 
	there is a unique node in the database to represent  $Elin$).
	\item An index is built on the \textit{symbol} of each atom, and a uniqueness constraint is defined on the couple \textit{symbol/terms} (implying, \eg, that there is a unique node in the database to represent atom $Authors(Elin, P_{269}$).
\end{enumerate}
\end{itemize}
The algorithms presented in the previous sections involve the construction of queries in Cypher to  be evaluated on  our  Neo4J database.
We now focus on two of them: one needed when chasing and one that computes the set \textsf{LinkedNull}.

\paragraph{Query for chasing.}
Chasing means applying constraints. The application of a constraint happens when its body can be instantiated by facts in the database.
Thus, to decide on the application of a constraint $c$, we need a query capable of :
\begin{enumerate}[label=(\arabic*)]
\item Verifying whether the database instance contains the facts necessary for the instantiation of $body(c)$ and
\item returning a non-empty answer only if a corresponding instantiation for $head(c)$ does not already exist in the database.
\end{enumerate}
In a logic formalism, if  $c$ is of the form
{$c:  L_1 (\alpha_1), \dots, L_m (\alpha_m) \rightarrow L_0 (\alpha_0)$}, 
we should write the query
{$q_{ch} : Q(\alpha) \leftarrow L_1 (\alpha_1), \dots, L_m (\alpha_m), not~L_0(\alpha_0)$}, where $\alpha$ is the list of variables corresponding to variables in $body(c)$, that is, variables universally quantified variables of $c$.
The idea here is: if $h_t$ is an instantiation such that  $h_t(body(c)) \subseteq \Db $, the query $q_{ch}$ has a non empty answer only if
$h'_t (L_0(\alpha_0))\not\in \Db$ for any extension $h'_t$ of $h_t$.

\begin{figure}
	\centering
		\begin{tikzpicture}[shorten >=3pt,->,node distance=3cm,on grid,every text node part/.style={align=center}]
		\node[draw,circle] (n6) {$n_6$\\Elin}; 
		\node[draw,above = 2cm of n6] (a27) {$n_{127}$\\Supervises}; 
		\node[draw,below = 2cm of n6] (a28) {$n_{128}$\\Supervises}; 
		\node[draw,circle,above = 2cm of a27] (n7) {$n_7$\\Sten}; 
		\node[draw,circle,left = of a28] (n8) {$n_8$\\Linda}; 
		\node[draw,above = 2cm of n7] (a29) {$n_{129}$\\Supervises}; 
		\node[draw,circle,above = 2cm of a29] (n10) {$n_{10}$\\Thor}; 
		\node[draw,right = of n6] (a2) {$n_{102}$\\Researcher}; 
		\node[draw,left = of n10] (a3) {$n_{103}$\\Researcher}; 
		\node[draw,left = of n7] (a4) {$n_{104}$\\Student}; 
		\node[draw,above = 2cm of n8] (a5) {$n_{105}$\\Student}; 
		\node[draw,right = of a28] (a16) {$n_{116}$\\Authors}; 
		\node[draw,right = of a27] (a17) {$n_{117}$\\Authors}; 
		\node[draw,color=red,right = of n10] (a18) {$n_{118}$\\Authors}; 
		\node[draw,circle,right = of a16] (n5) {$n_5$\\$P_{240}$}; 
		\node[draw,circle,right = of a17] (n9) {$n_9$\\$P_{269}$}; 
		\node[draw,circle,color=blue,fill=gray,right = of a18] (n11) {$n_{11}$\\$\_N_1$}; 
		\node[draw,above = 2cm of n5] (a21) {$n_{121}$\\Cites}; 
		\node[draw,color=red,below = 2cm of n11] (a24) {$n_{124}$\\Cites}; 
		\node[draw,circle,color=blue,below = 2cm of a24] (n12) {$n_{12}$\\$\_N_2$}; 
		\node[draw,color=red,below = 2cm of a18] (a11) {$n_{111}$\\Publication}; 
		\node[draw,color=red,left = of n12] (a12) {$n_{112}$\\Publication}; 
		
		\path
		(a2) edge [] node[above] {$r_2$} node[below] {rank: 0} (n6) 
		(a3) edge [] node[above] {$r_3$} node[below] {rank: 0} (n10) 
		(a4) edge [] node[above] {$r_4$} node[below] {rank: 0} (n7) 
		(a5) edge [] node[left] {$r_5$} node[right] {rank: 0} (n8) 
		(a11) edge [] node[above left, near start] {$r_{11}$} node[below right, near start] {rank: 0} (n11) 
		(a12) edge [] node[above] {$r_{12}$} node[below] {rank: 0} (n12) 
		(a16) edge [] node[below left, near start] {$r_{17}$} node[above right, near start] {rank: 0} (n6) 
		edge [] node[above] {$r_{18}$} node[below] {rank: 1} (n5) 
		(a17) edge [] node[above left, near start] {$r_{19}$} node[below right, near start] {rank: 0} (n6) 
		edge [] node[above] {$r_{20}$} node[below] {rank: 1} (n9) 
		(a18) edge [] node[above] {$r_{21}$} node[below] {rank: 0} (n10) 
		edge [] node[above] {$r_{22}$} node[below] {rank: 1} (n11) 
		(a21) edge [] node[left] {$r_{27}$} node[right] {rank: 0} (n9) 
		edge [] node[left] {$r_{28}$} node[right] {rank: 1} (n5) 
		(a24) edge [] node[left] {$r_{33}$} node[right] {rank: 0} (n11) 
		edge [] node[left] {$r_{34}$} node[right] {rank: 1} (n12) 
		(a27) edge [] node[right] {$r_{39}$} node[left] {rank: 0} (n6) 
		edge [] node[right] {$r_{40}$} node[left] {rank: 1} (n7) 
		(a28) edge [] node[right] {$r_{41}$} node[left] {rank: 0} (n6) 
		edge [] node[above] {$r_{42}$} node[below] {rank: 1} (n8) 
		(a29) edge [] node[right] {$r_{43}$} node[left] {rank: 0} (n10) 
		edge [] node[right] {$r_{44}$} node[left] {rank: 1} (n7) 
		;
		\end{tikzpicture}
	\caption{Graph database instance (extract). Optimization labels and attributes are omitted. \label{fig:runex-graph}}
\end{figure}
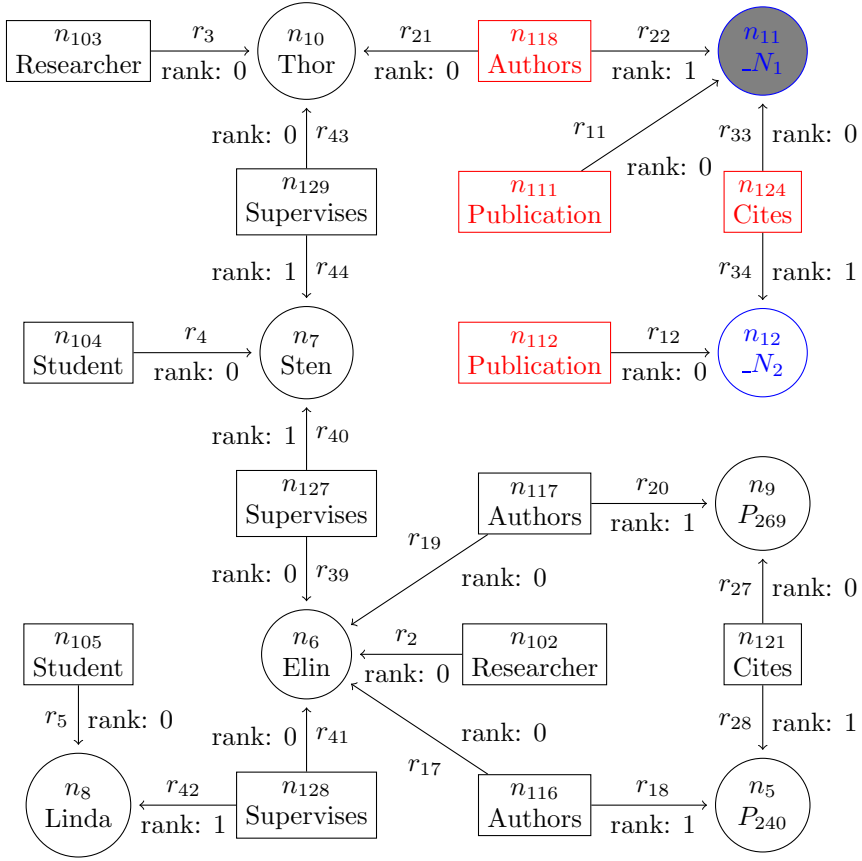

\begin{figure}
	\begin{lstlisting}[mathescape]
MATCH ($x_k$:Element {value: $t_i$})
MATCH ($a_1$:Atom {symbol: `$L_1$`}), ..., ($a_m$:Atom {symbol: `$L_m$`})
WHERE expr1 and NOT EXISTS { $\label{chase:not-exists}$
      MATCH ($a$:Atom { symbol: `$L_0$`})
      WHERE expr2
} RETURN {$\alpha|_1:x_1, \alpha|_2:x_2, \dots, \alpha|_k:x_k$}}
	\end{lstlisting}
	\caption{Cypher template for chasing \label{query-chase}}
\end{figure}

Figure~\ref{query-chase} shows  the Cypher template of query $q_{ch}$.
We first look for atoms that match $body(c)$.
	On the line~\ref{chase:not-exists} in Figure~\ref{query-chase}, the \verb|WHERE NOT EXISTS| clause is used to check that no instance of the $head(c)$ exists.
	Two expressions are built (\verb|expr1| and \verb|expr2|). 
	Terms in $\alpha$ are treated orderly. 
	Notice that \verb|expr1| is built for dealing with atoms in $body(c)$ and \verb|expr2| is built for dealing with the atom in $head(c)$.
	The first \verb|MATCH| acts as a starting point of the graph traversal. It is built with constants or nulls (\eg, \texttt{($x_k$}\texttt{:Element:Constant \{symbol:$t_i$\})}) as we usually consider constraints instantiated by insertions.
	Then the pattern,  built with the second \verb|MATCH| and the \verb|WHERE| clause, links the constants to the positions in the atoms of the  body.

	Separating the two MATCH allows us to guide the query planner to first search the constants (called node seeking) and then look for the connected nodes to find the atoms. This is important because, in doing so we drastically reduce the search space, because constants are unique values retrieved in $O(1)$, and only  \textbf{:Atom} connected nodes are searched, thus avoiding to visit all instance nodes of the predicate.
	
\begin{example}\label{ex:chase-ex}
{\rm
Considering the insertion of $Authors(Bob, P_1)$  in the database instance

\smallskip
$\Db= \{Supervises(Alice, Bob),$ $Authors(Alice, P_1) \}$

\smallskip\noindent
with the only constraint $c_6$ defined by:

\smallskip
$c_6 : Authors(X, P), Authors(Y, P), Supervises(X,Y) \rightarrow PhDPaper (Y, P,Z)$

\smallskip\noindent
Two instantiations $h$ and $h'$ should be checked: one on the first atom $Authors$ ($h(X) = Bob$, $h(P)= P_1$) and one for the second atom $Authors$
($h'(Y)=  Bob$, $h'(P) = P_1$).
 Figure~\ref{fig:qchase} shows the chase query for the  instantiation $h'$ of the constraint \ref{c6}. \hfill$\Box$}

\begin{figure}
	\begin{lstlisting}
MATCH (x0:Element:Constant {value: 'Bob'}),
      (x1:Element:Constant {value: 'P_1'})
MATCH (a0:Atom {symbol: 'Authors'  }),
      (a1:Atom {symbol: 'Authors'  }),
      (a2:Atom {symbol: 'Supervise'})
WHERE (a0)-[:Authors   {rank: 0}]->(x2),
      (a0)-[:Authors   {rank: 1}]->(x1),
      (a1)-[:Authors   {rank: 0}]->(x0),
      (a1)-[:Authors   {rank: 1}]->(x1),
      (a2)-[:Supervise {rank: 0}]->(x2),
      (a2)-[:Supervise {rank: 1}]->(x0)
and NOT EXISTS {
      MATCH (a:Atom {symbol: 'PhDPaper'})
      WHERE (a)-[:PhDPaper {rank: 0}]->(x0),
            (a)-[:PhDPaper {rank: 1}]->(x1)
} RETURN {`X`: x2.value, `Y`: x0.value, `P`: x1.value} AS sub
	\end{lstlisting}
	\caption{Cypher template for Example~\ref{ex:chase-ex} \label{fig:qchase}}
	\end{figure}
\end{example}

\paragraph{Query to find \textsf{LinkedNull} sets.}
Figure~\ref{fig:qPartition} presents the Cypher query that implements   the \textsf{LinkedNull} definition (Section~\ref{sec:simplification}) for building partitions of atoms. 
The clause  \verb|UNWIND| can transform any list  into individual rows.
For instance, if we consider a list \verb|['Elin', 'Sten']| of constant symbols, the clause  \verb|UNWIND| over such a list gives
a table with  one column $c$ and two rows whose values are  \verb|'Elin'| and \verb|'Sten'|.
In Figure~\ref{fig:qPartition}, the clause \verb|UNWIND| (line~\ref{linestart1}) is used to set nulls from a given list to our initial table
with one row for each null.
The goal of the first \verb|MATCH| (line~\ref{linematch1}) is to  select sub-graphs with atoms sharing the same null. 
On the line~\ref{linerange}, the range of the relationship (\verb|*1..|) indicates that node \verb|nullValueNode| can be connected to a node \verb|endNode| by a path
\verb|pathP| of arbitrary length.
Moreover, the direction is $\leftrightarrow$ indicates that \verb|pathP| can be composed by edges having any orientation.

The \verb|MATCH| clause looks for paths  starting with the  null of the  \verb|nullValueNode| to any other node representing an atom which is not \verb|nullValueNode| itself
(condition imposed by the \verb|WHERE| clause).
On the line~\ref{lineWith1}, the \verb|WITH| clause performs a `group by'.
 It allows to structure our working table with tuples where each null \verb|nullValueNode| is associated to a list of \verb|endNode|s (the nodes reached by paths \verb|pathP|).
On the line~\ref{lineWith2} a new organisation is built: \verb|linkedNodes| is divided into two lists, one containing nodes 
that represent predicate symbols  (\verb|linkedAtoms|)  and one for those representing nulls (\verb|linkedNulls|).
Notice that we place the initial node \verb|nullValueNode| in the  first position of the latter.
The resulting table partitions the atoms:  each atom  is associated to a list of nulls (those it is concerned by).
In the worst case, the former list contains all atoms having a null in the database.

\begin{figure}[htb]
	\begin{lstlisting}[escapechar=|]
UNWIND $nulls AS nullPredName |\label{linestart1}|
MATCH (nullValueNode:Element:Null {value: nullPredName}), |\label{linematch1}|
      pathP = (nullValueNode)-[*1..maxPathLength]-(endNode) |\label{linerange}|
WHERE endNode <> nullValueNode AND
      ALL(n IN nodes(pathP) WHERE NOT (n:Constant))
WITH COLLECT(DISTINCT endNode) AS linkedNodes, nullValueNode |\label{lineWith1}|
WITH |\label{lineWith2}|
[n IN linkedNodes WHERE (n:Atom)] AS linkedAtoms,
[nullValueNode] + [n IN linkedNodes WHERE (n:Null)] AS linkedNulls
UNWIND linkedAtoms AS a
RETURN a.symbol as a, a.terms as e, linkedNulls
	\end{lstlisting}
	\caption{Cypher template to find \textsf{LinkedNull} sets \label{fig:qPartition}}
\end{figure}

\begin{example}
\rm{
Considering the graph of Figure~\ref{fig:runex-graph},
if we search for atoms whose nulls are linked to $\_N_1$, \ie \verb|$nulls = ['_N1']|,  after the first MATCH in Figure~\ref{fig:qPartition},  we have:
	\begin{center}
		\begin{tabular}{ccl}
			\hline
			nullValueNode & endNode   & pathP                                                        \\
			\hline
			$n_{11}$ & $n_{111}$ & $[n_{11}, r_{11}, n_{111}]$                                  \\
			$n_{11}$ & $n_{118}$ & $[n_{11}, r_{22}, n_{118}]$                                  \\
			$n_{11}$ & $n_{124}$ & $[n_{11}, r_{33}, n_{124}]$                                  \\
			$n_{11}$ & $n_{12}$  & $[n_{11}, r_{33}, n_{124}, r_{34}, n_{12}]$                  \\
			$n_{11}$ & $n_{112}$ & $[n_{11}, r_{33}, n_{124}, r_{34}, n_{12}, r_{12}, n_{112}]$ \\
			\hline
		\end{tabular}
	\end{center}
	
	\noindent
	After the first WITH line~\ref{lineWith1}, we have:
	\begin{center}
		\begin{tabular}{ccc}
			\hline
			nullValueNode & linkedNodes                               \\
			\hline
			$n_{11}$ & $[n_{111}, n_{118}, n_{124}, n_{12}, n_{112}]$ \\
			\hline
		\end{tabular}
	\end{center}
	
	\noindent
	After the second WITH line~\ref{lineWith2}, we have:
	\begin{center}
		\begin{tabular}{ccc}
			\hline
			linkedAtoms                       & linkedNulls          \\
			\hline
			$[n_{111}, n_{118}, n_{124}, n_{112}]$ & $[n_{11}, n_{12}]$  \\
			\hline 
		\end{tabular}
	\end{center}
	\hfill$\Box$}
\end{example}

\subsection{Relational Data  Model}
\label{sec:query:sql}

Given an instantiated atom $P(t_1, \dots, t_n)$ in the logical representation oinf a database, our relational model consists in defining 
a table whose schema is $R_P[A_1, \dots A_n]$ where  all attributes are of type text. 
Notice that  $P(t_1, \dots, t_n)$ represents a  tuple on $R_P$ and, thus,
$(t_1, \dots, t_n)$ are values that  can be  constants or  nulls (nulls  have the  symbol  $\_$ as a prefix).
The translation of logical queries into SQL is straightforward.
However, some operations require the construction of procedures  to implement recursive queries.
Algorithm~\ref{findPartition} shows the  implementation of \textsf{LinkedNull} in the relational context.

We argue in this respect  that implementing Algorithm~\ref{findPartition} using a recursive SQL query is not efficient. Indeed, to do so
an additional table for storing the pairs of linked nulls is needed, and the following steps are necessary:
(a) a recursive SQL query to compute the transitive closure and
(b) a scan of the whole database to retrieve all corresponding atoms.
Moreover, the additional table needs to be maintained up to date after each update, which requires further processing.

{\small
	\algsetup{indent=1.5em}
	
	\begin{algorithm}[]
		\caption{FindLinkedNull($\Db, NullBucket)$ \label{findPartition}}
		\begin{algorithmic}[1]
%
			
			\STATE $newNull := \emptyset$, $allNulls := \emptyset$, $linkedNullSet := \emptyset$
			\WHILE{$NullBucket \neq \emptyset$}
			\STATE $allNulls := allNulls \cup NullBucket$
			\FORALL { table $R_P$  in the database schema}
			\FORALL{ tuple $u$  in $(select\;\; *\;\; from \;\;R_P$ $ where\;\; (A_1\;\; in \;\;NullBucket)\;\; or\;\; \dots\;\newline
			\hskip\algorithmicindent\;or\;\; (A_n\;\; in\;\; NullBucket) )$}
			\STATE  build atom $P(u)$;  add $P(u)$ in $linkedNullSet$
			\FORALL {null value $\_N \in null(u)$}
			\IF{$\_N \not \in allNulls$} 
			\STATE{$\;\;add \;\;\_N\;\; in\;\; newNull$}
			\ENDIF
			\ENDFOR
			\ENDFOR
			\ENDFOR
			\STATE $NullBucket := newNull$
			\STATE $ newNull := \emptyset$
			\ENDWHILE
			\RETURN $linkedNullSet$  
		\end{algorithmic}
	\end{algorithm}
}

We also notice that the implementation of 
 an incremental chase in the relational model follows the idea of setting up query $q_{ch}$ (as explained in Section~\ref{sec:graph}) which can be written as an SQL query involving a \verb|NOT EXISTS| clause.

\subsection{Discussion}
Querying graph database is significantly impacted by graph schema design.
The schema we have chosen transforms nulls into first-citizen elements and facilitates operations where,  by 'picking' a null,  we can easily detect all atoms connected (directly or indirectly)  to it. 
For instance, in Figure~\ref{fig:runex-graph}, if we 'pick' the null $\_N_1$ (the gray node $n_{11}$), we detect the atoms connected to it  together with other nulls (\ie $\_N_2$, the  blue node $n_{12}$).
In other words, this model  optimizes queries looking for linked nulls.
However, it may not be appropriate for other kinds of queries. 
For instance,  in the chase query, our model generates complex patterns that can be costly.
The relational model is less flexible than graph models, and thus its impact on querying is weaker.
However, relational model is not appropriate for the implementation of  recursion,  and nulls cannot be set as first-citizen element (identical null values appear repeatedly in the database instance).
Algorithm~\ref{findPartition} shows that to  implement \textsf{LinkedNull} we have to check null values for \textit{each} table, compromising the idea of an incremental approach.
On the other hand, the graph model is well suited for implementing incremental algorithms, because as seen in Section~\ref{sec:graph}, this model allows implementing \textsf{LinkedNull}  by visiting \textit{only} the atoms linked to nulls in $NullBucket$, as expected when considering an incremental computation.


\section{Experimental Results}
\label{sec:expe}

We gauge the performance of our incremental updating approach by analysing experiment results over a benchmark working on a graph (Neo4J) and a relational (MySQL) DBMS.
A benchmark run executes an update  on a database instance.

To build our database instances, we firstly view the original data sets from a FOL point of view.
Roughly speaking, a node or a relationship in the original data sets corresponds to a predicate  symbol, while their properties are the terms.
The conversion to our database models is straightforward, as presented in Section~\ref{sec:Impl}.
Nulls are inferred from already missing properties.
Constraints are hand-crafted, created from data observation and added to the databases we use for experiments.
The following three data sets are the basis of our instances:

\begin{itemize}[leftmargin=*]
\item \textit{Movie}\footnote{\url{https://github.com/neo4j-graph-examples/movies}}, available as a Neo4J instance, is a collection of data concerning movies, actors, directors. This data set contains  7  predicate symbols (with arity 2-4).

\item \textit{GOT}\footnote{\url{https://github.com/neo4j-graph-examples/graph-data-science}},  available as a Neo4J instance, deals with  the interactions between different characters in the book \textit{Game of Throne}. This data set contains 19 predicate symbols  (artiy 2-14).

\item \textit{LDBC}\footnote{\url{https://ldbcouncil.org/benchmarks/graphalytics/}}, available as a data set of the Linked Data Benchmark Council, offers synthetic data sets  for benchmarking. This data set contains 23  predicate symbols (artiy 1-2).
\end{itemize}
From the LDBC  data sets we build several instances, by  varying their size or the number of nulls.
To control the size of instances, their construction is  the result of:
\begin{enumerate*}[label=(\roman*)]
    \item  randomly selecting $k$ facts,  respecting the  distribution of the original data set  and, then,
    \item applying the \num{39}  hand-made constraints  on them.
\end{enumerate*}
The result is  a consistent  database instance with nulls.
Figure~\ref{table:xp} presents a summary of our database instances (or samples).
It is worth noting that, for example, an instance denoted as  LDBC 1K,  comes from a random selection of
\num{1000} facts  which  evolves to  \num{2248}  after the chase and core processing.
To control the number of nulls, we proceed as follows:
we take the largest LDBC instance, \ie with  \num{10000} facts,  and replace all nulls with constants.
Then, we choose, randomly, some constants that are replaced by linked nulls.
Figure~\ref{table:xp} presents  database instances  used in our runs, eight having nulls,  and one non-null instance.
All the database instances  are generated just once.
By following this creation process,  they  are consistent and minimal.

\begin{figure}[htb]
    \centering
    \begin{tabular}{l|c|c|c|c}
        Database       & Nb of facts & Nb of nulls & Nb of rules & Null/Facts  ($\tau$) \\
        \hline
        \hline
        Movie          & 604         & 340         & 12          & 0.56                 \\
        GameOfThrone   & 24818       & 17232       & 32          & 0.69                 \\
        \hline
        LDBC 1K        & 2248        & 190         & 39          & 0.08                 \\
        LDBC 10K       & 16559       & 1183        & 39          & 0.07                 \\
        \hline
        LDBC 10K 0N    & 16559       & 0           & 39          & 0.00                 \\
        LDBC 10K 50N   & 16559       & 50          & 39          & 0.00                 \\
        LDBC 10K 100N  & 16559       & 100         & 39          & 0.01                 \\
        LDBC 10K 500N  & 16559       & 500         & 39          & 0.03                 \\
        LDBC 10K 1000N & 16559       & 1000        & 39          & 0.06                 \\
    \end{tabular}
    \caption{Database instances (our  samples). \label{table:xp}}
\end{figure}

Runs are built from instances in Figure~\ref{table:xp} by
\begin{enumerate*}[label=(\roman*)]
    \item varying the update type (insertion or deletion);
    \item altering the size of the update (\numlist{1;5;10;20} atoms)  and
    \item augmenting artificially the number of facts in an instance. This latter step is done through the duplication of data
\end{enumerate*}
$n$-times (\numlist{1;2;5}), together with the renaming of the constants and the null names at each copy.

Each  run  performs \num{10} iterations plus \num{3} warm-up iterations (\ie  an ordinary iteration used to preload the system and database cache) not counted in the execution time.
Between each iteration,  the original database instance is restored, and
the Java garbage collector is triggered for consistent time measuring.
The benchmarks are implemented  in Java 16 with MySQL 8 and Neo4J 4.1  and
executed on a Rocky Linux 8.7 virtual server with \num{4} vCPU and \SI{16}{\giga\byte} of memory (\num{8} reserved for the database and \num{5} for the Java program) through docker 20.10.21.
In the docker container of a database instance, the  average of read/write on  disk is \SI{1}{\giga\byte\per\second}.
The same server hosts:
\begin{enumerate*}[label=(\roman*)]
    \item  one database server at the time and
    \item the benchmarks with only \num{4} vCPU.
\end{enumerate*}

Notice that, even if this configuration allows us to assess our implementations over different DBMS, our experiment performances are not representative of real world situations, where more powerful and dedicated hardware is available.

%

We first compare the incremental approach presented in this paper to the from-scratch in-memory approach in~\cite{CHL20}. For this aspect,  comparisons are performed only on the database \textit{Movie} because
the from-scratch in-memory version requires  a huge amount of memory for its computation.
We have  an average of \SI{9017}{\ms} for an update with the in-memory version and scale of 1 (initial size of the instance). 
MySQL has an average of \SI{151}{\ms} and Neo4J has \SI{2380}{\ms}.
For this small instance, the incremental approach is comparable with the from-scratch approach.
Considering an instance five times larger, we get an average of \SI{888966}{\ms} for the in-memory version, \SI{595}{\ms} for MySQL and \SI{2706}{\ms} for Neo4J.
Thus, it should be clear that using a DBMS in which an incremental version of update processing is implemented, allows for efficiently updating large databases that do not fit in main memory.

Next, we analyse the performance of  incremental updating  \wrt\
the number of atoms (database size) and nulls of an instance.
We denote by \textit{incompleteness degree} the number of distinct \textsf{LinkedNull} sets on a database.
We also investigate the number of \textit{queries} generated to interact with the DBMS.
Figure~\ref{xp:plots} presents  our experiment results.
On each plot,  the right axis, indicates the total number of facts in the instance. 
The curves show the average of resulting values for all runs corresponding to the displayed
abscissa.

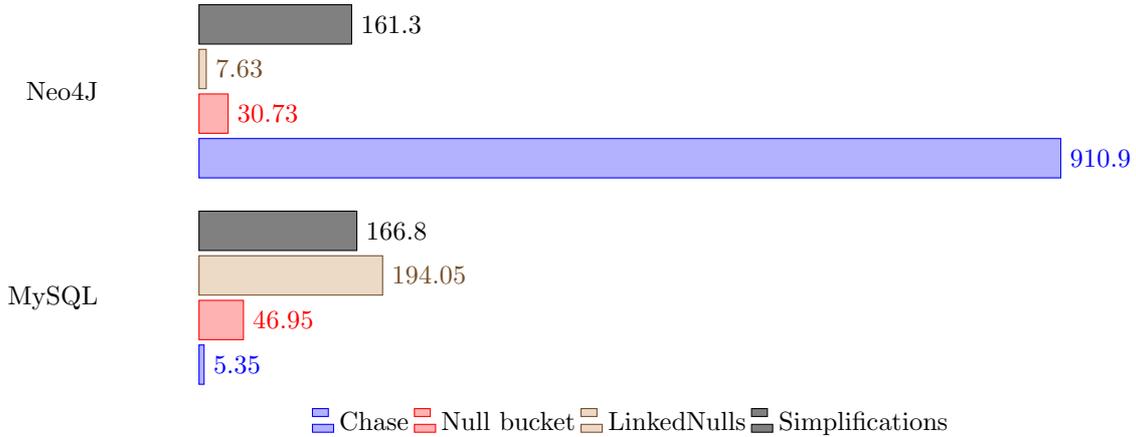
\begin{figure}
    \centering
    \begin{adjustbox}{width=\linewidth}
        \begin{tikzpicture}
            \pgfplotsset{scaled x ticks=false, scaled y ticks=false}
            \begin{axis}[xbar, axis x line=none, axis line style={draw=none}, tick style={draw=none}, symbolic y coords = {MySQL,Neo4J,Neo4J (rels typing)}, enlarge y limits = 0.5, restrict x to domain=0:*, nodes near coords, ytick=data, xlabel={Time}, x unit=\si{\ms}, legend columns=-1, legend style={draw=none, at={(0.5,0)}, anchor=north}, width=\columnwidth, height=.35\textheight, bar width=1.5em]
                \addplot table [y=db, x=chase, col sep=comma] {time_per_db.csv};
                \addlegendentry{Chase};

                \addplot table [y=db, x=nullBucket, col sep=comma] {time_per_db.csv};
                \addlegendentry{Null bucket};

                \addplot table [y=db, x=partitions, col sep=comma] {time_per_db.csv};
                \addlegendentry{LinkedNulls};

                \addplot table [y=db, x=simplifications, col sep=comma] {time_per_db.csv};
                \addlegendentry{Simplifications};
            \end{axis}
        \end{tikzpicture}
    \end{adjustbox}
    \caption{Average time of each operation per DBMS (\si{\ms}) removing outsiders with more than \SI{30}{\s} differences }
    \label{xp:time:db}
\end{figure}

We first note that the update type (insertion or deletion) has no real impact on the  performance of our approach.
Figure~\ref{xp:query:simple} shows that the number of queries is linear on the number of nulls, except for three  down spikes when the degree of incompleteness of the database instance is low.
This is the case for the database \textit{Movies}, and the down spikes coincide to a situation where only this database is concerned.
Indeed, thanks to the use of multiple data sets, we observe here that the predicate arity (\ie the number of edges per node or the number of columns in a table) may have an impact on  our results.
Linearity \wrt\ the  number of nulls is explained by the fact that
consistency preservation implies the generation of new data linked by their nulls.
Thus, due to our construction method, bigger databases imply more linked nulls (\ie bigger \textsf{LinkedNull} sets).
Incremental updates generate $q_{Bucket}$ queries  to retrieve impacted nulls.
Bigger databases likely have more impacted nulls willing to be simplified during the core computation, increasing the number of necessary $q_{core}$ queries.

Consequences of bigger \textsf{LinkedNull} sets are:
\begin{enumerate}[label=(\roman*)]
    \item in MySQL, Algorithm~\ref{findPartition}  generates a large amount of queries and
    \item in Neo4J,  the unique query  needed to retrieve a \textsf{LinkedNull} set is more complex and, thus, more time-consuming.
\end{enumerate}
However,  this augmentation is negligible  as our model  is designed to optimize such a  query (Figure~\ref{xp:time:full}).


Experimental results in terms of execution time of our updating approach is shown in Figures~\ref{xp:time:simple:mysql} and~\ref{xp:time:simple:neo}.
In MySQL (Figure~\ref{xp:time:simple:mysql}),  update execution time is  linear in the number of nulls while the database size has little impact.
Indeed, as  the number of queries increases with the number of nulls, update execution time  in MySQL increases accordingly.
In Neo4J  (Figure~\ref{xp:time:simple:neo}), update  execution time is more significantly impacted by the size of the instance.

The explanation of this discrepancy comes from the separate analysis of  the performance of the  main operations of our approach (Figures~\ref{xp:time:db} and~\ref{xp:time:full}).
The data model chosen in the Neo4J version optimizes the retrieval of \textsf{LinkedNull}, but is not appropriate to operations involving simplification
(Section~\ref{sec:simplification}). Such operations involve complex pattern matching which are known to be expensive.
The chase (Figure~\ref{xp:time:db}) is the most expensive operation for Neo4J, mainly due to the  fact that it  includes a  simplification step
(\eg, if $A(a,b) \in \Db$ and $A(a, N_1)$  is generated by a constraint, then the insertion of $A(a, N_1)$ is canceled).

For the sake of readability, plots do not show results on \textit{GOT} instances with more than \num{17000} nulls. The results on this data set are similar: execution  time  evolves  linearly \wrt\ nulls in  MySQL and follows the size of the database in Neo4J.
With the \textit{GOT} runs, we achieve a mean execution time of \SI{14634}{\ms} with MySQL and \SI{5216}{\ms} with Neo4J for \num{24818} facts and \num{17232} nulls.
Increasing the size to \num{124090} facts and \num{86160} nulls rises run time to \SI{156132}{\ms} with MySQL and to \SI{203140}{\ms} with Neo4J.

\begin{figure}
    \centering
    \footnotesize
    \pgfplotsset{scaled x ticks=false, scaled y ticks=false}
    \begin{subfigure}{\linewidth}
        \begin{adjustbox}{width=\linewidth}
            \begin{tikzpicture}
                \begin{axis}[enlarge x limits = 0, ymin=0, xlabel={Nb of nulls}, ylabel={Nb of queries}, legend columns=-1, legend style={draw=none, at={(0.4,-0.15)}, anchor=north}, width=\textwidth, height=.4\textheight]
                    \addplot table [x=nulls, y=mysqlAdd, col sep=comma] {query_per_nulls.csv};
                    \addlegendentry{MySQL (INS)};

                    \addplot table [x=nulls, y=mysqlDel, col sep=comma] {query_per_nulls.csv};
                    \addlegendentry{MySQL (DEL)};

                    \addplot table [x=nulls, y=neoAdd, col sep=comma] {query_per_nulls.csv};
                    \addlegendentry{Neo4J (INS)};

                    \addplot table [x=nulls, y=neoDel, col sep=comma] {query_per_nulls.csv};
                    \addlegendentry{Neo4J (DEL)};
                \end{axis}
                \begin{axis}[enlarge x limits = 0, ymin=0, hide x axis, ylabel near ticks, yticklabel pos=right, ylabel={Nb of facts}, legend style={draw=none, at={(0.89,-0.15)}, anchor=north}, width=\textwidth, height=.4\textheight]
                    \addplot[no marks, dashed] table [x=nulls, y=facts, col sep=comma] {query_per_nulls.csv};
                    \addlegendentry{Number of facts};
                \end{axis}
            \end{tikzpicture}
        \end{adjustbox}
        \caption{Number of queries per null}
        \label{xp:query:simple}
    \end{subfigure}
    \medskip
    \begin{subfigure}{\linewidth}
        \begin{adjustbox}{width=\linewidth}
            \begin{tikzpicture}
                \begin{axis}[enlarge x limits = 0, ymin=0, xlabel={Nb of nulls}, ylabel={Time}, y unit=\si{\ms}, legend columns=-1, legend style={draw=none, at={(0.4,-0.15)}, anchor=north}, width=\textwidth, height=.4\textheight]
                    \addplot table [x=nulls, y=mysqlAdd, col sep=comma] {time_per_nulls.csv};
                    \addlegendentry{Insert};

                    \addplot table [x=nulls, y=mysqlDel, col sep=comma] {time_per_nulls.csv};
                    \addlegendentry{Delete};
                \end{axis}
                \begin{axis}[enlarge x limits = 0, ymin=0, hide x axis, ylabel near ticks, yticklabel pos=right, ylabel={Nb of facts}, legend style={draw=none, at={(0.63,-0.15)}, anchor=north}, width=\textwidth, height=.4\textheight]
                    \addplot[no marks, dashed] table [x=nulls, y=facts, col sep=comma] {time_per_nulls.csv};
                    \addlegendentry{Number of facts};
                \end{axis}
            \end{tikzpicture}
        \end{adjustbox}
        \caption{Time per null for MySQL}
        \label{xp:time:simple:mysql}
    \end{subfigure}
    \caption{Benchmarks results of 540 scenarios, average over 10 runs}
\end{figure}
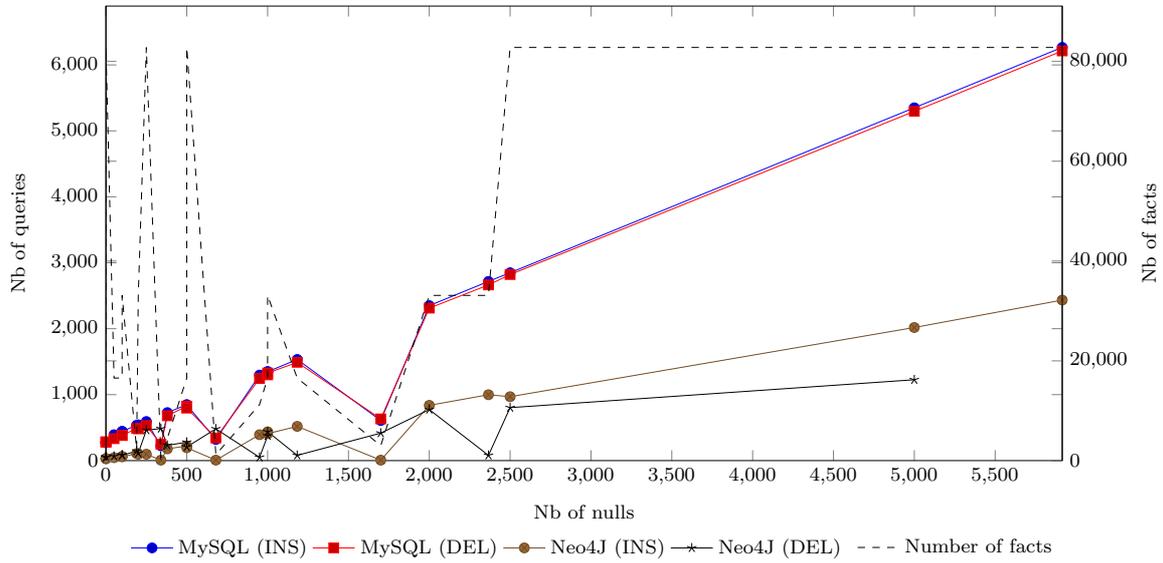
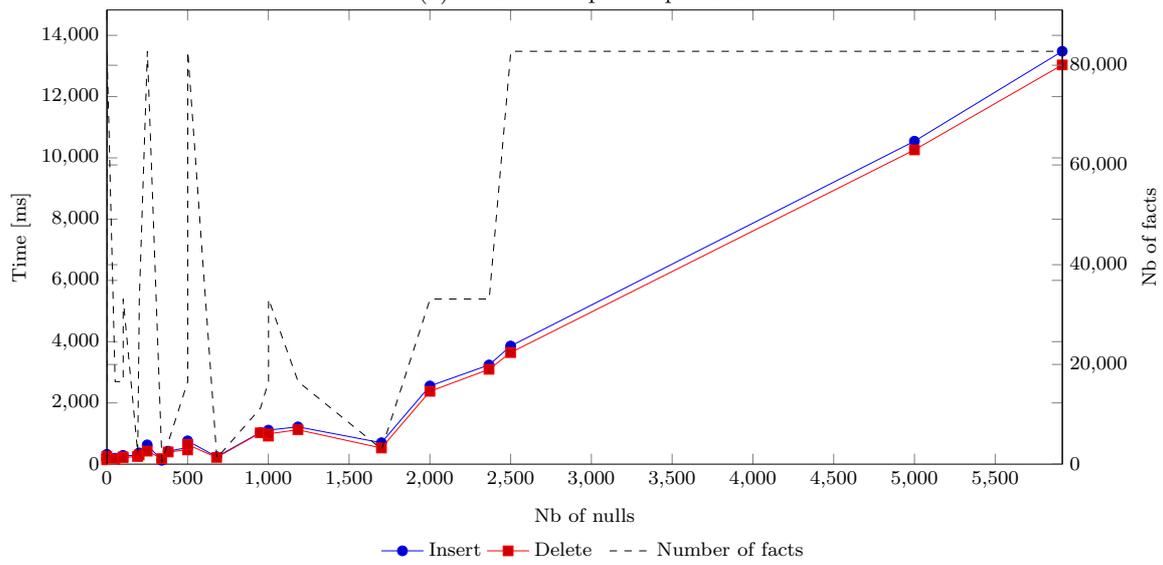
\begin{figure}
    \ContinuedFloat
    \centering
    \footnotesize
    \pgfplotsset{scaled x ticks=false, scaled y ticks=false}
    \begin{subfigure}{\linewidth}
        \begin{adjustbox}{width=\linewidth}
            \begin{tikzpicture}
                \begin{axis}[enlarge x limits = 0, ymin=0, xlabel={Nb of nulls}, ylabel={Time}, y unit=\si{\ms}, legend columns=-1, legend style={draw=none, at={(0.4,-0.15)}, anchor=north}, width=\textwidth, height=.4\textheight]
                    \addplot table [x=nulls, y=neoAdd, col sep=comma] {time_per_nulls.csv};
                    \addlegendentry{Insert};

                    \addplot table [x=nulls, y=neoDel, col sep=comma] {time_per_nulls.csv};
                    \addlegendentry{Delete};
                \end{axis}
                \begin{axis}[enlarge x limits = 0, ymin=0, hide x axis, ylabel near ticks, yticklabel pos=right, ylabel={Nb of facts}, legend style={draw=none, at={(0.63,-0.15)}, anchor=north}, width=\textwidth, height=.4\textheight]
                    \addplot[no marks, dashed] table [x=nulls, y=facts, col sep=comma] {time_per_nulls.csv};
                    \addlegendentry{Number of facts};
                \end{axis}
            \end{tikzpicture}
        \end{adjustbox}
        \caption{Time per null for Neo4J}
        \label{xp:time:simple:neo}
    \end{subfigure}
    \medskip
    \begin{subfigure}{\linewidth}
        \begin{adjustbox}{width=\linewidth}
            \begin{tikzpicture}
                \begin{axis}[enlarge x limits = 0, ymin=0, xlabel={Nb of nulls}, ylabel={Time}, y unit=\si{\ms}, legend columns=4, legend style={draw=none, at={(0.5,-0.15)}, anchor=north}, width=\textwidth, height=.4\textheight]
                    \addplot table [x=nulls, y=mysqlChase, col sep=comma] {time_per_nulls_details.csv};
                    \addlegendentry{MySQL (chase)};

                    \addplot table [x=nulls, y=mysqlNullBucket, col sep=comma] {time_per_nulls_details.csv};
                    \addlegendentry{MySQL (null bucket)};

                    \addplot table [x=nulls, y=mysqlPartitions, col sep=comma] {time_per_nulls_details.csv};
                    \addlegendentry{MySQL (linked nulls)};

                    \addplot table [x=nulls, y=mysqlSimplifications, col sep=comma] {time_per_nulls_details.csv};
                    \addlegendentry{MySQL (simplifications)};

                    \addplot table [x=nulls, y=neoChase, col sep=comma] {time_per_nulls_details.csv};
                    \addlegendentry{Neo4J (chase)};

                    \addplot table [x=nulls, y=neoNullBucket, col sep=comma] {time_per_nulls_details.csv};
                    \addlegendentry{Neo4J (null bucket)};

                    \addplot table [x=nulls, y=neoPartitions, col sep=comma] {time_per_nulls_details.csv};
                    \addlegendentry{Neo4J (linked nulls)};

                    \addplot table [x=nulls, y=neoSimplifications, col sep=comma] {time_per_nulls_details.csv};
                    \addlegendentry{Neo4J (simplifications)};
                \end{axis}
                \begin{axis}[enlarge x limits = 0, ymin=0, hide x axis, ylabel near ticks, yticklabel pos=right, ylabel={Nb of facts}, legend style={draw=none, at={(0.5,-0.3)}, anchor=north}, width=\textwidth, height=.4\textheight]
                    \addplot[no marks, dashed] table [x=nulls, y=facts, col sep=comma] {time_per_nulls_details.csv};
                    \addlegendentry{Number of facts};
                \end{axis}
            \end{tikzpicture}
        \end{adjustbox}
        \caption{Time of operations per null}
        \label{xp:time:full}
    \end{subfigure}
    \caption[]{Benchmarks results of 540 scenarios, average over 10 runs}
    \label{xp:plots}
\end{figure}

\paragraph{Reproducibility.}
Results obtained by our experiments are reproducible through the use of the benchmarks and implementation available in \url{https://gitlab.com/jacques-chabin/UpdateChase}.

\section{Related Works}
\label{sec:RW}

Our work goals include   four important features of modern applications: incompleteness, consistency as a measure of quality, incremental tools for  efficient data processing and adaptability to graph data models.

Solid basis have been established for treating incompleteness of  relational databases 
~\cite{FKUV86,Gra91,IL84,Rei86,Zan84}, particularly for querying.
Much less attention has been given to updates on incomplete databases, although important work, such as~\cite{AbG85,FUV83,Win90} can be cited.
Today, integrating and exchanging data are very common,  leading to the proliferation of applications involving \textit{dynamic incomplete data}
on emerging  data models that deal with more general graph-structured data. Incompleteness beyond the relational data model has received much less attention~\cite{Sir14}, and, in this context, updating with respect to constraints  is rarely considered. 
Indeed,  consistency maintenance  is usually left aside in favour of efficiency, which can prove costly when we are concerned with the quality of analytical results.
Work such as~\cite{HLS98,LiS02,ScT98} witnesses the complexity of the problem of keeping a database consistent with respect to constraints in a dynamic environment.
In~\cite{FKAC13,HHU17,HaL17} we find newer proposals, adapted to the RDF world,  that considers constraints  in our traditional database viewpoint (\ie not in the web semantic  standard way, where constraints are  just inference rules~\cite{GOP11,LMS08,PaS15}).
It is worth noting that the use of  tuple generating constraints (TGD)   increases expressiveness at the cost of difficulties
that involve  a chase procedure 
(cf. a survey in~\cite{Onet13}, a benchmark in~\cite{BKM17}) 
 to compute semantics
and 
the  generation of side effects in an update context - imposing extra insertions or deletions (\wrt\ those required by the user)  to preserve  consistency.
The literature offers sufficient conditions to avoid a non-terminating chase which consist in limiting the format of constraints.
We instead introduce $\delta_{max}$, keeping the possibility of dealing with any kind of constraints while avoiding infinite processing.
Furthermore, we use  simplifications to keep the database instance as small as possible and to avoid the presence of useless nulls, \ie
database maintenance consists in keeping its \textit{core} (which follows the ideas in~\cite{FKP05})  whose implementation is ensured by a simplification routine performed 
in association to update routines. 

In brief, data analytic tools become essential in different application domains and  their quality relies on data consistency.
But in order to deal with huge scale applications, we must aim at efficient data processing solutions~\cite{Sir14}, bringing incremental solutions to the front of the stage, particularly when working with new data models (as done in the XML context~\cite{ABHLM04,BPV04,BH03}).
In the context of graph databases, the approach in \cite{FTX21} proposes a method for `incrementalizing' graph algorithms abstracted in a fix-point model.
Our approach cannot be summarized by that proposal. As seen before, we can outline our method in the expression
$\Db' = core_{|NullBucket}(upd_{|U}((\Db\diamondsuit U))$ where $U$ is the set of user's required updates - this set is \textit{increased} through an inference process that generates side-effects.  The proposal in~\cite{FTX21} needs a 'complete' set of updates as input. 
In other words, our fix-point operation involves changes on the update set while in~\cite{FTX21} the update set is fixed. Their goal is to incrementally compute new answers on an updated graph and not to incrementally update the graph.
As the \textit{core} computation is not a fix-point one, it is not in the scope of~\cite{FTX21}.

Finally, our experiments reinforce the idea that graph schema design has a significant impact on query performance. 
Our graph schema is  designed to optimize one type of query and performs badly to those that differ widely.
Schema  optimization may be a solution:  as  in~\cite{ALQEO21}, in this paper,  it is done through techniques that reduce edge transversal.


\section{Conclusions}
\label{sec:conclusion}

This paper contributes to improve the maintenance of consistent incomplete databases by proposing incremental routines that interact with database systems. It extends prior work  in~\cite{CHL20} where a \textit{from-scratch}  in-memory method was proposed. 
Two  implementations of our approach, one under a graph database model and one under the traditional relational database model, are presented. 
Experiment results raise questions about the representation of nulls in a graph database.
Indeed, this work is also a step towards incremental updating attributed graphs with incomplete data. 
It illustrates the impact of schema graph design in querying and, consequently, in the performance of an incremental updating approach that relies on two main queries: 
one that looks for linked nulls and another that looks for redundant atoms willing to be simplified. 
Property graph model has an increasingly important role today,  the handling of nulls in such a model is related to schema definition and query optimization issues that need to be further explored.


\bibliographystyle{acm}
\bibliography{references}

\end{document}